\documentclass[preprint,resetfootnote]{aastex701}

\journalinfo{\ }

\begin{document}

\title{A Processing Workflow for Cassini VIMS Jupiter Cubes}

\author[orcid=0000-0002-9305-1901,gname=Asier,sname=Anguiano-Arteaga]{Asier Anguiano-Arteaga}
\affiliation{Departamento de F\'{\i}sica Aplicada, Escuela de Ingenier\'{\i}a de Bilbao, Universidad del Pa\'{\i}s Vasco/Euskal Herriko Unibertsitatea (UPV/EHU), Bilbao, Spain}
\affiliation{Department of Physics, Atmospheric, Oceanic and Planetary Physics, University of Oxford, Oxford, UK}
\email[show]{asier.anguiano@ehu.eus}
\correspondingauthor{Asier Anguiano-Arteaga}

\author[gname=Patrick G. J.,sname=Irwin]{Patrick G.~J. Irwin}
\affiliation{Department of Physics, Atmospheric, Oceanic and Planetary Physics, University of Oxford, Oxford, UK}
\email{patrick.irwin@physics.ox.ac.uk}

\author[gname=Santiago,sname=P\'{e}rez-Hoyos]{Santiago P\'{e}rez-Hoyos}
\affiliation{Departamento de F\'{\i}sica Aplicada, Escuela de Ingenier\'{\i}a de Bilbao, Universidad del Pa\'{\i}s Vasco/Euskal Herriko Unibertsitatea (UPV/EHU), Bilbao, Spain}
\email{santiago.perez@ehu.eus}

\author[gname=Davide,sname=Grassi]{Davide Grassi}
\affiliation{Istituto di Astrofisica e Planetologia Spaziali--Istituto Nazionale di Astrofisica (IAPS-INAF), Roma, Italy}
\affiliation{Max-Planck-Institut f\"ur Sonnensystemforschung (MPS),  G\"ottingen, Germany}
\email{davide.grassi@inaf.it}

\author[gname=Emiliano,sname=D'Aversa]{Emiliano D'Aversa}
\affiliation{Istituto di Astrofisica e Planetologia Spaziali--Istituto Nazionale di Astrofisica (IAPS-INAF), Roma, Italy}
\email{emiliano.daversa@inaf.it}

\begin{abstract}

We present a calibrated catalog of Cassini Visible and Infrared Mapping Spectrometer (VIMS) observations of Jupiter, together with the processing workflow used to generate the final publicly available products. Starting from the raw VIMS cubes, the workflow produces radiometrically consistent multi-extension FITS files and includes a revised visible-channel calibration, a revised infrared-channel calibration that resolves a subset of problematic cases not satisfactorily treated by the standard ISIS pipeline, corrections for pointing-related misalignments between spectral cubes and geometric backplanes, and customized dark signal correction strategies. The final products include calibrated spectral cubes together with geometry backplanes and wavelength information for subsequent scientific analysis. We assess the consistency of the calibrated products through internal validation tests and comparisons with independent reference spectra from the literature. The resulting products provide a uniform and validated data set of Cassini VIMS Jupiter observations for community use. The full catalog is available as a public \dataset[data set]{https://doi.org/10.5281/zenodo.19223781}.

\end{abstract}



\section{Introduction}
\label{introduction}

The Cassini spacecraft was launched on October 15, 1997, as part of the Cassini--Huygens mission to the Saturn system. Whereas Cassini was intended to investigate Saturn, its rings, and its satellites from orbit, the Huygens probe was designed to enter the atmosphere of Titan and study it in situ. On its interplanetary trajectory, Cassini performed a gravity-assist flyby of Jupiter in late 2000 and early 2001 before reaching Saturn in 2004 \citep{Matson2002}. During the Jupiter encounter, Cassini’s instruments obtained a valuable set of observations and measurements of Jupiter under a wide range of geometries and observing conditions. These spanned both the approach phase, at relatively low solar phase angles, and the flyby phase, which included closest approach and rapidly changing conditions, with some observations extending across the terminator into the nightside.

Onboard Cassini, the Visible and Infrared Mapping Spectrometer (VIMS) was an imaging spectrometer covering the approximate spectral range from 0.35 to 5.1~$\mu$m through two partially overlapping channels: a visible channel (VIS), spanning about 0.35--1.05~$\mu$m, and an infrared channel (IR), spanning about 0.9--5.1~$\mu$m \citep{Brown2004}. The VIS channel samples the spectrum at about 7.3~nm over 96 bands, whereas the IR channel samples it over 256 bands with a mean spectral sampling of about 16.6~nm \citep{McCord2004}. The two channels differ not only in spectral coverage but also in the way the image cube is acquired. In the VIS channel, radiation is focused onto a spectrometer slit, dispersed by a diffraction grating, and imaged onto a CCD matrix detector, so that spatial information along the slit and spectral information in the dispersion direction are recorded simultaneously; the second spatial dimension is then built up by scan-mirror motion. In the IR channel, by contrast, a linear detector records the spectrum of a single spatial element, and the two-dimensional scene is assembled sequentially by scanning the field of view combining the motion of two internal mirrors. These two acquisition schemes are commonly referred to as pushbroom (VIS) and whiskbroom (IR), respectively \citep{Coradini2004}.
In nominal spatial mode, both channels provide an effective instantaneous field of view (IFOV) of about
$0.5~\mathrm{mrad} \times 0.5~\mathrm{mrad}$ per spatial pixel,
while the VIS channel also supports a high-resolution mode reaching about
$0.17~\mathrm{mrad} \times 0.17~\mathrm{mrad}$ per spatial pixel
\citep{McCord2004}.
For the VIS channel, these IFOV values refer to the angular dimensions of an individual spatial sampling element along the slit, rather than to the full slit itself.
The IR channel also includes a one-direction high-resolution mode, yielding a rectangular IFOV of about
$0.5~\mathrm{mrad} \times 0.25~\mathrm{mrad}$ per spatial pixel,
although no observations from the Jupiter encounter are available in that mode.

Cassini/VIMS observations from the Jupiter encounter have supported a variety of scientific applications. Early VIS-channel analyses used VIMS data to study the spatial variability of Jupiter's visible spectrum across the disk, with particular emphasis on methane and ammonia absorption features \citep{Coradini2004}. Subsequent studies used VIMS observations to investigate Jupiter's emitted power in combination with Cassini/CIRS \citep{Li2012}, tropospheric cloud structure from the 5~$\mu$m thermal-emission spectrum \citep{Giles2015}, nightside H$_3^+$ emission \citep{Stallard2015}, and the spectral properties of jovian aerosols and chromophores across belts, zones, and the Great Red Spot \citep{Sromovsky2017,Baines2019}. More recently, disk-integrated VIMS spectra of Jupiter have been used as analogs for extrasolar giant planets and brown dwarfs \citep{Coulter2022}.

Despite these previous applications, a homogeneous, fully documented, and publicly available set of calibrated Jupiter products has remained desirable. There are several reasons for reprocessing the Jupiter cubes. First, in the visible channel, \citet{Sromovsky2017} found that the original VIMS calibration implemented in the Integrated Software for Imagers and Spectrometers (ISIS, \citealt{Anderson2004}) of the United States Geological Survey (USGS) appeared to require an upward correction of about 10\% and presented their disk-averaged VIMS reference spectrum scaled by a factor of 1.12. As shown below, the VIS calibration adopted in this work naturally incorporates this scaling through the responsivity-based implementation itself, without the need for any additional empirical rescaling. Second, in the IR channel, some approach-phase observations calibrated with the default ISIS pipeline produce clearly over-scaled reflected spectra, with values exceeding physically plausible levels because of a failure mode in the default Sun--target distance normalization. Third, the Jupiter data set reveals pointing-related mismatches between VIS radiometric cubes and geometric backplanes, as well as channel-dependent dark signal issues that are not optimally handled by the default ISIS pipeline. Finally, a dedicated reprocessing makes it possible to deliver public and reproducible calibrated data products in a uniform format suitable for subsequent scientific analysis.

The workflow presented here starts from the raw VIMS archive files and produces calibrated multi-extension files in the Flexible Image Transport System (FITS) format \citep{Pence2010}. The final products include calibrated VIS and IR spectral cubes, wavelength vectors, full width at half maximum (FWHM) vectors, and geometry backplanes in a common format designed for direct scientific use. A summary of the delivered products is provided in Appendix~\ref{appendix:summary_products}. The processing uses ISIS together with SPICE geometry information \citep{Acton1996, Acton2017}, but the final radiometric calibration itself is implemented through custom VIS and IR procedures tailored to the Jupiter data set analyzed here. The complete data set, together with the responsivity matrices used in the calibration, is publicly available \citep{AnguianoArteaga2026Zenodo}.

This paper is organized as follows. Section~\ref{sec:initial_data_processing} describes the selection of the raw VIMS Jupiter observations and the ISIS-based processing. Section~\ref{sec:radiometric_calibration} presents the radiometric calibration adopted for the VIS and IR channels. Section~\ref{sec:cube_wavelengths} discusses the wavelength assignment of the final cubes. Section~\ref{sec:calibration_validation} validates the calibrated products through comparisons with independent reference spectra from the literature. Section~\ref{sec:geom_backplanes} describes the geometry backplanes and the treatment of residual pointing-related issues. Appendix~\ref{appendix:summary_products} summarizes the final delivered products, and Appendix~\ref{appendix:dark_correction} describes the dark signal and stripe-correction strategies adopted in the VIS and IR channels.

\section{Initial Data Set and ISIS-Based Processing}
\label{sec:initial_data_processing}

\subsection{Selection of the Raw VIMS Data}

The raw Cassini VIMS observations analyzed in this work were retrieved using the Outer Planets Unified Search (OPUS) archive search tool\footnote{\url{https://opus.pds-rings.seti.org/}} \citep{French2021}. We restricted the search to observations of Jupiter acquired between 2000 November 16 and 2001 January 11. 

Additional selection criteria were then applied to identify products suitable for the present analysis. In practice, observations with an observation duration shorter than 21~s often did not show a recognizable Jupiter disk and instead consisted of uniform black-and-white frames or a pixelated multicolor ``checkerboard''-like pattern. We therefore adopted a minimum observation duration of 21~s.

A further useful constraint was the OPUS quantity \textit{Greater Size in Pixels}, defined for a two-dimensional observation as the number of pixels along the longer of the two image axes. We imposed a minimum value of 14. When no lower limit was applied, the search again returned many clearly non-informative products, dominated by uniform black-and-white frames or pixelated multicolor ``checkerboard''-like patterns. In addition, some of the returned observations contained only a very small portion of Jupiter's disk in the frame, with the disk sampled at very low spatial resolution. Applying a minimum \textit{Greater Size in Pixels} value of 14 allowed us to discard nearly 1,500 unusable or very low-quality cubes.

The \textit{Is Prime} archive flag was set to \textit{No} in the OPUS query. Although this flag identifies whether VIMS was the prime instrument for the observation (as opposed to a ride-along while another instrument was prime), including cases with \textit{Is Prime} = \textit{Yes} would have added only 24 extra products, 8 of which displayed distorted planetary disks. All of them were nominal spatial-resolution cubes acquired during the approach phase and did not add useful coverage beyond that already provided by the selected data set.

For the VIS channel, both nominal- and high-resolution observations were considered. No equivalent restriction was imposed for the IR channel, but in practice all IR products retained in the final data set were acquired in nominal mode. Some IR cubes were acquired with the archive \textit{Sampling mode} flag set to \textit{Under}. This corresponds to a non-nominal IR spatial sampling configuration that is not supported by the standard ISIS camera model; consequently, \texttt{spiceinit} fails for these cubes and their geometry information cannot be recovered. These IR cubes were therefore excluded from the final data set. The simultaneously acquired VIS cubes with the same time stamp, by contrast, can still be processed. The initial archive search returned 474 VIS cubes and 474 IR cubes. Of the IR cubes, 151 (31.9\%) were acquired in \textit{Sampling mode} = \textit{Under}, leaving 323 IR cubes after their exclusion.

\subsection{ISIS-Based Processing}
Part of the processing described in this work used version 8.3.0 of the USGS Integrated Software for Imagers and Spectrometers\footnote{\url{https://isis.astrogeology.usgs.gov/}}
 (ISIS; \citealt{Anderson2004}). Within this framework, the original VIMS archive products were converted into ISIS \texttt{.cub} files, SPICE initialization \citep{Acton1996} was performed to provide the pointing and geometric information required by the camera model, and geometry backplanes were generated.

The radiometric calibration adopted in this work was not carried out with \texttt{vimscal}, the standard ISIS application for calibrating VIMS cubes. Instead, the final VIS and IR calibrations were implemented through custom procedures tailored to the Jupiter data set analyzed here. The \texttt{vimscal} outputs were nevertheless used as a reference for comparison. Additional ISIS inputs and products used in the workflow included label metadata such as detector offsets and exposure times, calibration files, geometry backplanes generated with the ISIS application \texttt{phocube}, and pointing corrections applied with the ISIS application \texttt{deltack}.

\section{Radiometric Calibration}
\label{sec:radiometric_calibration}

Radiometric calibration converts raw VIMS data from detector units (digital numbers, DN) into instrument-independent physical units. The final product adopted here is the dimensionless quantity $I/F$. This quantity, widely used in planetary remote sensing, normalizes the measured radiance by the incident solar irradiance, thereby removing the $1/d^2$ dependence of the solar flux at the target and facilitating comparisons between observations acquired at different Sun--target distances and illumination geometries. The quantity $I/F$ is defined as

\begin{equation}
I/F (\lambda) = \frac{\pi I(\lambda)}{F_{\odot}(\lambda,d)}
= \frac{\pi d^2 I(\lambda)}{F_{\odot}(\lambda,1~\mathrm{AU})},
\end{equation}

where $I(\lambda)$ is the measured spectral radiance, in units of $\mathrm{W\,m^{-2}\,sr^{-1}\,\mu m^{-1}}$, and $F_{\odot}(\lambda,d)=F_{\odot}(\lambda,1~\mathrm{AU})/d^2$ is the solar spectral irradiance at the target distance $d$, with $d$ expressed in AU and $F_{\odot}$ in units of $\mathrm{W\,m^{-2}\,\mu m^{-1}}$. Here $F_{\odot}(\lambda,1~\mathrm{AU})$ denotes the corresponding solar spectral irradiance at 1 AU. Radiance can be recovered straightforwardly from $I/F$ through

\begin{equation}
I(\lambda) = I/F(\lambda)\;\frac{F_{\odot}(\lambda,d)}{\pi}
= I/F(\lambda)\;\frac{F_{\odot}(\lambda,1~\mathrm{AU})}{\pi d^2}.
\end{equation}

\begin{figure}
\centering
\includegraphics[width=0.5\columnwidth]{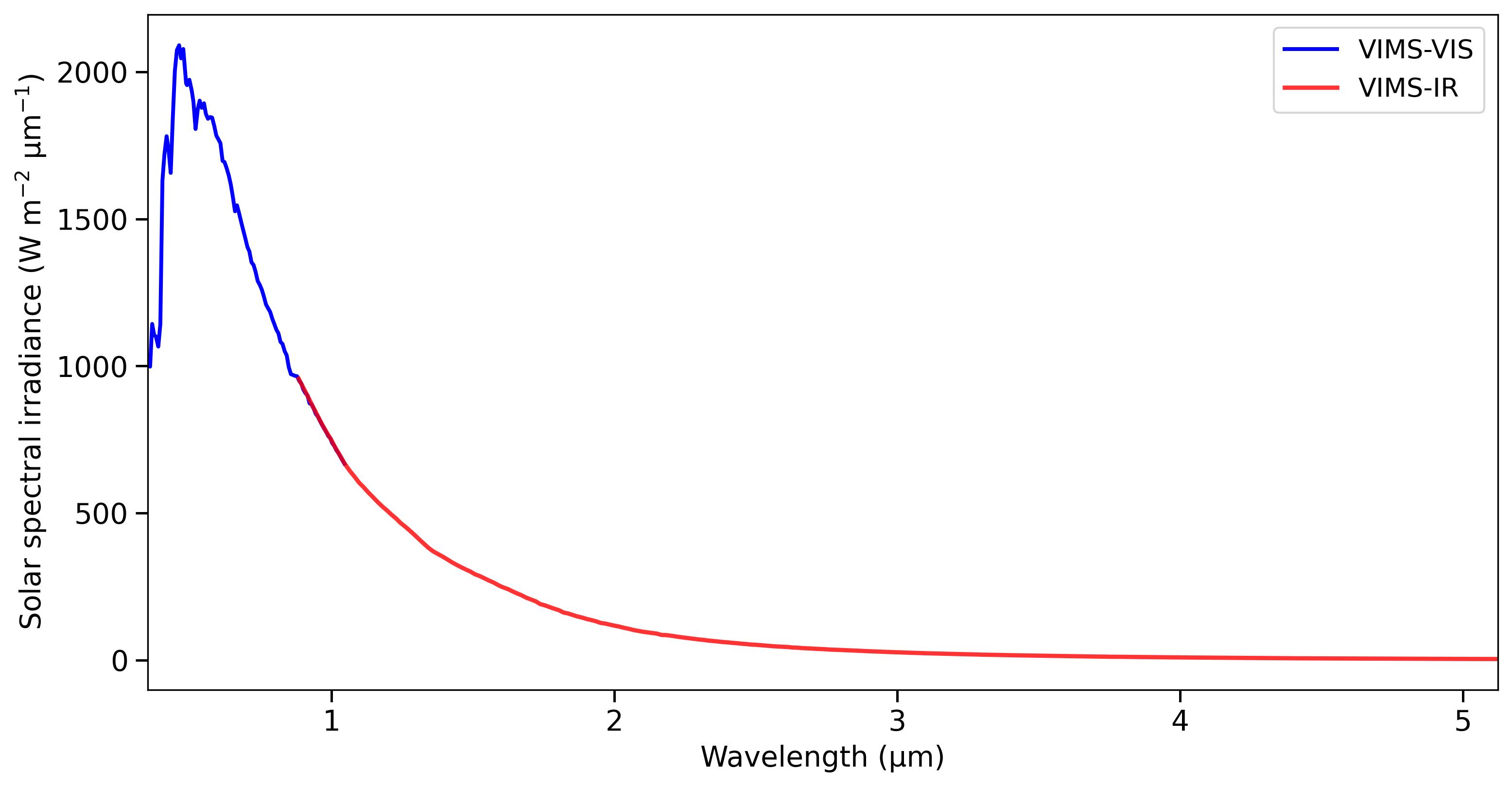}
\caption{Solar spectral irradiance from the ISIS VIMS solar cube \texttt{VIMS2000.9610.solar\_v0003.cub}, corresponding to the latest calibration set available for the year 2000. The cube provides the high-resolution solar spectrum of \citet{Thompson2015} convolved to the VIMS-VIS (blue) and VIMS-IR (red) bandpasses.}
\label{fig:solar_vims}
\end{figure}

Figure~\ref{fig:solar_vims} shows the solar spectral irradiance contained in the ISIS solar cube \texttt{VIMS2000.9610.solar\_v0003.cub}, corresponding to the latest calibration set available for the year 2000. This product provides a solar spectrum already convolved to the VIMS-VIS and VIMS-IR spectral responses. The underlying high-resolution solar reference is given by \citet{Thompson2015}.

\subsection{VIS Channel}
\label{sec:VIScal}

\subsubsection{General Processing and Calibration Pipeline}

For the visible channel (0.35--1.05~$\mu$m), we reproduce, following \citet{Filacchione2006}, the radiometric calibration implemented at INAF--IAPS (the ``Rome pipeline''), using the same two-dimensional responsivity matrices and detector geometry.

\paragraph{Input data.}
The input to the VIS calibration consists of raw VIS cubes in ISIS \texttt{.cub} format (see Appendix~\ref{appendix:dark_correction} for the dark signal treatment). From the ISIS label we read the sampling mode (NOMINAL or HI-RES), the detector sample offset \texttt{X\_OFFSET}, the scan-mirror offset \texttt{Z\_OFFSET}, the visible-channel exposure time $t_{\mathrm{exp}}$ (in seconds), and the observation start time and target name, which are used to compute the Sun--Jupiter distance $d$ in astronomical units (AU). All VIS data used here were acquired in low gain and do not require any gain-related scaling.

The VIS exposure time $t_{\mathrm{exp}}$ is obtained from the \texttt{EXPOSUREDURATION} keyword, explicitly selecting the VIS value when multiple numbers are present and converting from milliseconds to seconds. The Sun--Jupiter distance $d$ is computed from the observation start time using JPL DE432s ephemerides via Astropy~\citep{Astropy}.

\paragraph{VIS responsivity matrices.}
The core of the calibration is a pair of two-dimensional responsivity matrices, stored as \texttt{float64} arrays and originally described in \citet{Filacchione2006}. 
This calibration used in-flight lunar observations to determine the absolute response of a reference CCD pixel, and Venus-flyby observations to derive the flat-field response across the detector. This differs from the nominal ISIS/\texttt{vimscal} calibration, which follows the standard VIMS calibration history based on pre-launch characterization and in-flight refinements using Venus, the Moon, Galilean satellites, and stars \citep{McCord2004}, with later updates included in the RC19 calibration of \citet{Clark2018}, the final VIMS wavelength and radiometric calibration release.
The adopted responsivity matrices are publicly available as part of the released data set \citep{AnguianoArteaga2026Zenodo}:
\begin{itemize}
    \item \texttt{VIMS\_V\_NOMINAL\_RESPONSIVITY\_2005\_25.DAT} (nominal mode),
    \item \texttt{VIMS\_V\_HIGHRES\_RESPONSIVITY\_2005\_25.DAT} (high-resolution mode).
\end{itemize}

Each file contains a two-dimensional array of size $[96, N_{\rm samples}]$, where 96 is the number of VIS spectral bands and $N_{\rm samples}$ is the full detector extent in the along-slit direction. The array is transposed to obtain a matrix $R_{\rm tot}(x,\lambda)$ with dimensions $[N_{\rm samples},96]$, giving the responsivity for each detector column and wavelength. Here $x$ denotes the detector-column (sample) index.

\paragraph{Exact detector slicing in the sample direction.}
The VIS DN cube is written as $\mathrm{dn}(x,z,\lambda)$, where $x$ is the detector sample coordinate and $z$ is the line (scan) index. From the ISIS label we obtain the sampling mode (NOMINAL or HI-RES) and the detector offset \texttt{X\_OFFSET}; we define
\begin{equation}
x_0 = \texttt{X\_OFFSET} - 1
\end{equation}
to convert the 1-based convention used in the labels to 0-based array indexing.

The responsivity is tabulated on a detector-sample grid of 192 columns in HI-RES mode (unbinned) and 64 columns in NOMINAL mode (corresponding to $3\times3$ spatial binning, i.e., $192/3$), whereas the acquired frames are smaller windows within these full coordinate systems. In practice, the processed VIS cubes used here have spatial sizes of $32\times32$ pixels in NOMINAL mode and $64\times64$ pixels in HI-RES mode.

In NOMINAL mode (IFOV = 0.5 mrad pixel$^{-1}$), \texttt{X\_OFFSET} specifies the first column of the detector window in the binned sample coordinate, so the relevant responsivity columns form a contiguous block starting at $x_0$, expressed in pixel units:
\begin{equation}
\mathrm{start} = x_0, \qquad \mathrm{stop} = x_0 + \mathrm{width}.
\end{equation}

In HI-RES mode (IFOV = 0.17 mrad pixel$^{-1}$), the detector window is defined on the unbinned 192-column sample coordinate and is referenced to the optical boresight, which corresponds to the center of the grid (sample 95 in 0-based arrays). The relevant detector columns are therefore selected as
\begin{equation}
\mathrm{start} = x_0 + 95 - \mathrm{width}/2, \qquad
\mathrm{stop} = x_0 + 95 + \mathrm{width}/2.
\end{equation}

We then extract
\begin{equation}
R_{\rm sub}(x,\lambda) = R_{\rm tot}[\mathrm{start}:\mathrm{stop},\lambda],
\end{equation}
which has dimensions $[\mathrm{width},96]$ and matches the spatial width of the cube.

The VIS responsivity matrices depend only on detector sample $x$ and wavelength $\lambda$. This is appropriate for the VIS channel because VIMS-V operates as a pushbroom imaging spectrometer: the full slit is recorded simultaneously on the detector, and successive image lines are generated only by scan-mirror motion. By contrast, the IR channel uses a whiskbroom acquisition scheme, in which the two-dimensional scene is built sequentially by two perpendicular scan motions. Consequently, the IR responsivity matrix is treated as a function of detector sample $x$, detector line $z$, and wavelength $\lambda$, so \texttt{Z\_OFFSET} is relevant only for the IR channel.

\paragraph{Conversion from DN to $I/F$.}
Once the appropriate sub-matrix $R_{\rm sub}(x,\lambda)$ has been obtained, the VIS reflectivity is computed as
\begin{equation}
\frac{I}{F}(x,z,\lambda) =
\frac{\mathrm{dn}(x,z,\lambda)\,R_{\rm sub}(x,\lambda)\,d^2}{t_{\mathrm{exp}}}.
\end{equation}

Here $\mathrm{dn}(x,z,\lambda)$ are the VIS counts in the ISIS cube, $R_{\rm sub}(x,\lambda)$ is the responsivity at detector column $x$ and wavelength $\lambda$, in units of $(I/F)/(DN\,s^{-1})$, $t_{\mathrm{exp}}$ is the VIS exposure time in seconds, and $d$ is the Sun--Jupiter distance in AU. This expression reproduces the VIS calibration described by \citet{Filacchione2006}.

\paragraph{Final VIS products.}
The calibrated VIS $I/F$ cubes are written as multi-extension FITS files. The primary Header/Data Unit (HDU) contains the spectral cube in $I/F$, and the additional HDUs store the geometry backplanes described in Section~\ref{sec:geom_backplanes}. The final two extensions contain the wavelengths and their corresponding FWHM values, respectively. For visualization, we adopt the standard astronomical convention in which the lower-left pixel has image coordinates $(1,1)$ and $y$ increases upward. The FITS cubes are therefore written in a display-oriented convention that can differ from the native detector orientation of the original samples--lines array, so that, when opened in commonly used FITS viewers such as SAOImage DS9 \citep{Joye2003} or QFitsView \citep{Ott2012}, they reproduce the intended view of Jupiter, with north up and south down.

\subsubsection{Sanity Checks}

\begin{figure}[ht!]
\centering
\includegraphics[width=0.65\columnwidth]{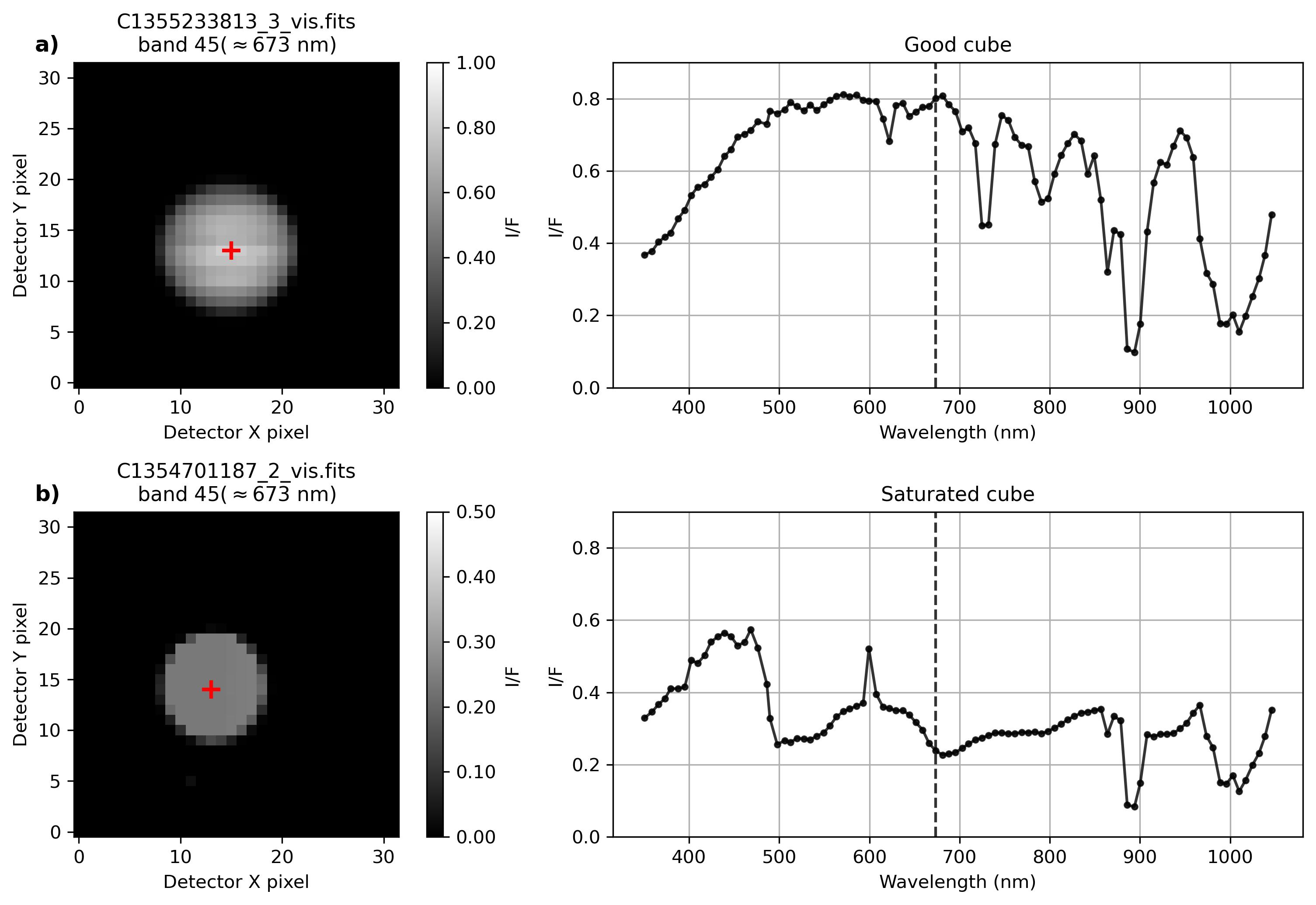}
\caption{Example VIMS-VIS cubes with (a) a well-behaved spectrum and (b) a saturated spectrum. The dashed vertical line indicates the band used for the image display.}
\label{fig:vis_saturation_examples}
\end{figure}

All VIS cubes were visually inspected to identify data-quality issues. For each cube, we examined images together with spectra extracted from the disk. A large number of cubes were found to be globally saturated over most of the VIS wavelength range (Figure~\ref{fig:vis_saturation_examples}).

Of the 474 VIS cubes initially downloaded, 86 were rejected during an initial screening step because they were corrupted or severely distorted, leaving 388 candidate cubes for the sanity-check stage. Of these, 326 cubes were discarded owing to saturation, leaving 62 well-behaved VIS cubes. The discarded cases result from acquisition-level issues and cannot be reliably corrected a posteriori within our processing pipeline.

Despite this substantial rejection rate, the retained VIS data set still provides good coverage of observing conditions and scene types. It spans a wide range of phase angles and includes observations reaching spatial resolutions of up to about 5,000 km pixel$^{-1}$ in nominal mode and up to about 1,700 km pixel$^{-1}$ in high-resolution mode. It also contains cases in which features such as the Great Red Spot (GRS) appear at different locations across the disk.

\subsection{IR Channel}
\subsubsection{General Processing and Calibration Pipeline}
\label{sec:ir_general_pipeline}

The processing of the VIMS infrared channel (0.88--5.12~$\mu$m) closely parallels the visible-channel procedure described in Section~\ref{sec:VIScal}, but uses IR-specific calibration files and responsivity matrices. 
The conversion to I/F follows the standard RC19 calibration files used by ISIS/\texttt{vimscal} \citep{Clark2018}; here, however, we implement it through explicitly archived responsivity matrices, allowing the calibration applied to the released Jupiter products to be reproduced directly. This treatment also allows us to avoid the problematic implicit Sun--target distance calculation in ISIS for approach-phase cubes, as discussed in Section~\ref{sec:validation_IRchannel}.

\paragraph{Input data.}
The input to the IR calibration consists of raw IR cubes in ISIS \texttt{.cub} format (see Appendix~\ref{appendix:dark_correction} for the dark signal treatment). From the ISIS label we read the sampling mode, the detector sample offset \texttt{X\_OFFSET}, the detector line offset \texttt{Z\_OFFSET}, the IR exposure time $t_{\mathrm{exp}}$ (in seconds), and the observation start time and target name, which are used to compute the Sun--Jupiter distance $d$ in astronomical units (AU). All IR data used here were acquired in low gain and do not require any gain-related scaling.

Following the ISIS \texttt{vimscal} procedure, we define an effective integration time as

\begin{equation}
t_{\mathrm{eff}} = t_{\mathrm{exp}} \times 1.01725 - 0.004~\mathrm{s},
\end{equation}

where the factor 1.01725 corrects the IR exposure duration for the known inaccuracy of the VIMS clock\footnote{\url{https://isis.astrogeology.usgs.gov/8.1.0/Application/presentation/Tabbed/vimscal/vimscal.html}}, and the subtraction of 0.004~s accounts for the settling time after each movement of the scanning secondary mirror \citep{Clark2018}. In practice, the exposure duration recorded in the label is given in milliseconds, so it is first converted to seconds before applying this correction. The Sun--Jupiter distance $d$ is obtained in the same way as for the VIS cubes, using the observation start time and JPL DE432s ephemerides via Astropy~\citep{Astropy}.

\paragraph{IR responsivity matrices.}
For the IR channel, we construct a global responsivity matrix $R_{\mathrm{tot}}(x,z,\lambda)$ from the RC19 radiometric calibration data set \texttt{VIMS2000.9610} \citep{Clark2018}. The construction mirrors the VIS case: we compute a band-dependent factor $K(\lambda)$ in units of $(I/F)/(\mathrm{DN}\,\mathrm{s}^{-1})$ using the RC19 multiplier (wavelength-dependent spectral response), the IR wavelength-calibration cube, and the solar spectrum at 1 AU, and combine it with the detector flatfield, taken from the ISIS calibration cube \texttt{ir\_flatfield\_v0002.cub}, to introduce the $x$- and $z$-dependence.

Unlike the VIS case, however, the IR flatfield is three-dimensional. Accordingly, the IR responsivity matrix is treated as a function of detector sample $x$, detector line $z$, and wavelength $\lambda$, such that

\begin{equation}
R_{\mathrm{tot}}(x,z,\lambda) = \frac{K(\lambda)}{\mathrm{FLAT}(x,z,\lambda)},
\end{equation}

with dimensions $[N_{\mathrm{samples}}, N_{\mathrm{lines}}, 256]$. This matrix encodes the wavelength and detector-position dependence of the IR response and is defined over the full detector extent, independently of the size of any individual cube. In computing $K(\lambda)$, we adopt the constant multiplicative factor $K_{\mathrm{RC19,IR}} = 8112.0$ as reported by \citet{Clark2018}.

For the Jupiter data set considered here, all retained IR observations were acquired in NOMINAL sampling mode. We therefore use the NOMINAL responsivity matrix listed below, stored as a \texttt{float32} array. This precision is sufficient for the $I/F$ conversion because the underlying ISIS calibration inputs are 32-bit data, while using \texttt{float64} would increase file size and read time without providing any practical improvement for the calibration. This matrix is publicly available as part of the released data set \citep{AnguianoArteaga2026Zenodo}.

\begin{itemize}
    \item \texttt{VIMS\_IR\_NOMINAL\_RESPONSIVITY\_2000\_9610.DAT} (nominal mode).
\end{itemize}

\paragraph{Exact detector slicing.}
For a given cube, the spatial dimensions of the detector window are \texttt{width} pixels in the detector-sample direction and \texttt{height} pixels in the detector-line direction. Using the detector offsets \texttt{X\_OFFSET} and \texttt{Z\_OFFSET}, we extract from the global responsivity matrix $R_{\mathrm{tot}}(x,z,\lambda)$ the subset corresponding to the field of view of the cube.

In NOMINAL sampling mode (IFOV = 0.5~mrad~pixel$^{-1}$), $R_{\mathrm{tot}}(x,z,\lambda)$ is tabulated on the nominal detector grid (64 samples and 64 lines in the flatfield used here), and the cube samples a contiguous window within this grid. We define

\begin{equation}
x_0 = \texttt{X\_OFFSET} - 1, \qquad z_0 = \texttt{Z\_OFFSET} - 1,
\end{equation}

to convert the 1-based label convention to 0-based array indexing, and use

\[
\mathrm{start}_x = x_0, \qquad \mathrm{stop}_x = x_0 + \mathrm{width},
\]
\begin{equation}
\mathrm{start}_z = z_0, \qquad \mathrm{stop}_z = z_0 + \mathrm{height}.
\end{equation}

The responsivity matrix actually used for the cube is then

\begin{equation}
R_{\mathrm{sub}}(x,z,\lambda) =
R_{\mathrm{tot}}[\mathrm{start}_x:\mathrm{stop}_x,\,
\mathrm{start}_z:\mathrm{stop}_z,\lambda],
\end{equation}

which has dimensions $[\mathrm{width}, \mathrm{height}, 256]$ and matches the spatial size and spectral sampling of the IR cube.

\paragraph{Conversion from DN to $I/F$.}
Given the ISIS DN cube $\mathrm{dn}(x,z,\lambda)$, the cropped responsivity matrix $R_{\mathrm{sub}}(x,z,\lambda)$, the effective integration time $t_{\mathrm{eff}}$, and the Sun--Jupiter distance $d$, the IR reflectivity is computed as

\begin{equation}
\frac{I}{F}(x,z,\lambda) =
\frac{\mathrm{dn}(x,z,\lambda)\,R_{\mathrm{sub}}(x,z,\lambda)\,d^2}
{t_{\mathrm{eff}}}.
\end{equation}

Here $\mathrm{dn}(x,z,\lambda)$ are the dark-corrected IR counts, $R_{\mathrm{sub}}(x,z,\lambda)$ is the responsivity at detector sample $x$, line $z$, and wavelength $\lambda$, $t_{\mathrm{eff}}$ is the effective integration time in seconds, and $d$ is the Sun--Jupiter distance in AU.

\paragraph{Final IR products.}
The calibrated IR $I/F$ cubes are written as multi-extension FITS files, following the same structure and visualization convention as the VIS products. The primary HDU contains the spectral cube in $I/F$, and the additional HDUs store the geometry backplanes described in Section~\ref{sec:geom_backplanes}. The final two extensions contain the wavelengths and their corresponding full width at half maximum (FWHM), respectively.

\subsubsection{Sanity Checks}

\begin{figure}[ht!]
\centering
\includegraphics[width=0.65\columnwidth]{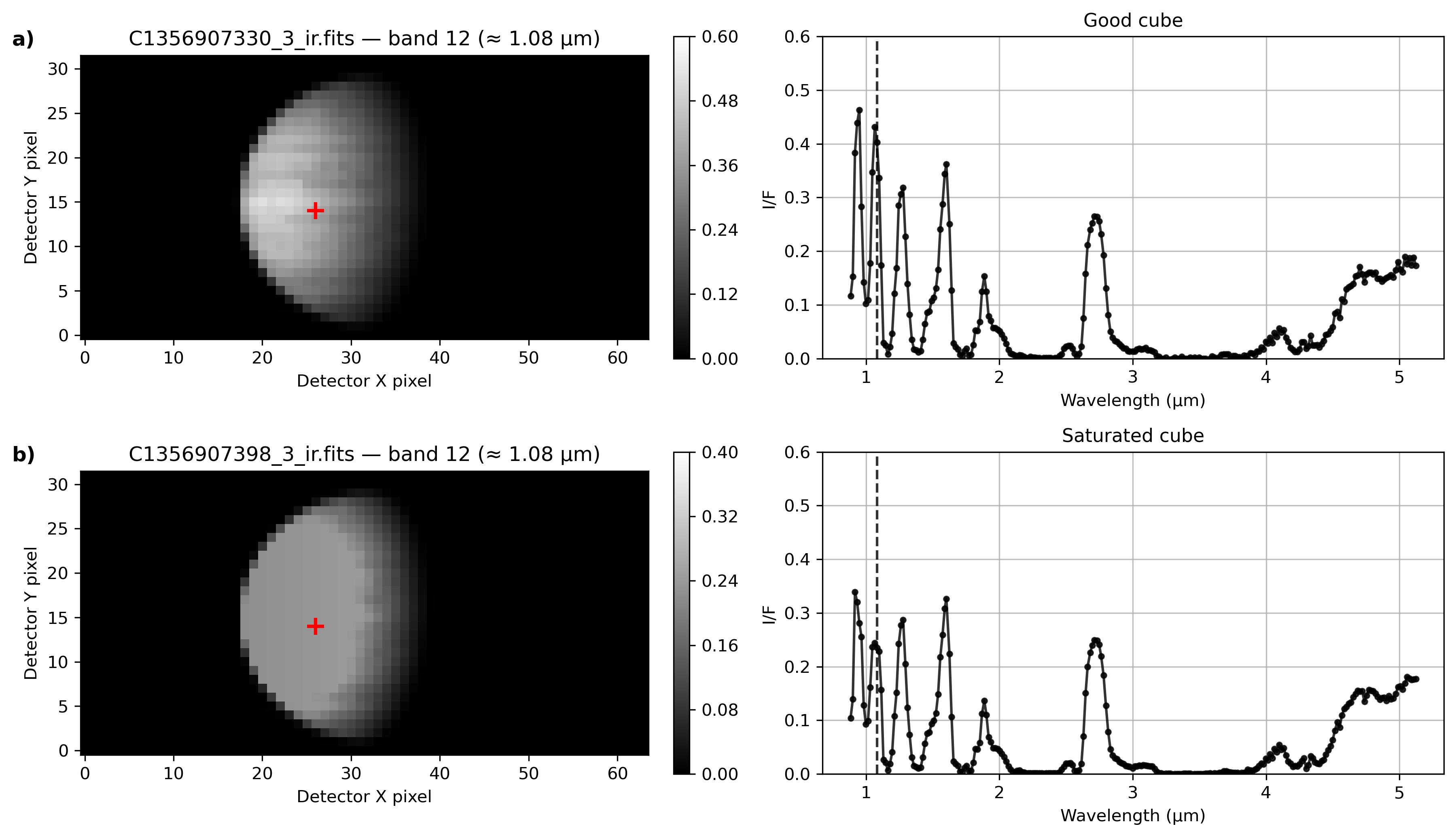}
\caption{Example VIMS-IR cubes with (a) a well-behaved spectrum and (b) a saturation-affected spectrum at shorter IR wavelengths. The dashed vertical line in each spectral panel indicates the band used for the image display.}
\label{fig:ir_saturation_example}
\end{figure}

All IR cubes were visually inspected to identify pathological spectra. For each cube, we examined images together with spectra extracted from the disk. As in the VIS case, a large number of cubes were found to be affected by saturation at the shorter IR wavelengths, leading to defective products. An example comparing a well-behaved cube with a saturation-affected cube is shown in Figure~\ref{fig:ir_saturation_example}.

The saturation-affected IR cubes were excluded from the final data set because the compromised signal cannot be recovered reliably after acquisition. Of the 474 IR cubes initially identified, 177 were rejected during an initial screening step because they were corrupted, severely distorted, or acquired in \textit{Sampling mode} = \textit{Under}, leaving 297 candidate cubes for the spectral quality-control stage. Of these, 226 cubes were discarded, leaving 71 well-behaved IR cubes.

Despite this substantial rejection rate, the retained data set still provides good coverage of observing conditions and scene types, including cases with smaller and larger apparent disk sizes, a broad range of phase angles, and observations in which features such as the Great Red Spot (GRS) appear at different locations across the disk.

\section{Cube Wavelengths}
\label{sec:cube_wavelengths}

The final multi-extension FITS products include, in their last two extensions, the wavelength sampling ($\lambda$) and the corresponding spectral bandpasses (FWHM) associated with each spectral channel. The reference standard wavelength set and FWHM values were taken from the RC19 compilation of \citet{Clark2018}.

For the VIS channel, we adopt the standard wavelengths without any time-dependent correction. \citet{Clark2018} explicitly state that the sampled wavelengths of the visible channel do not shift with time, and that the ground-calibrated wavelength scale is therefore adequate for the VIS channel.

\begin{figure}[ht!]
\centering
\includegraphics[width=0.6\columnwidth]{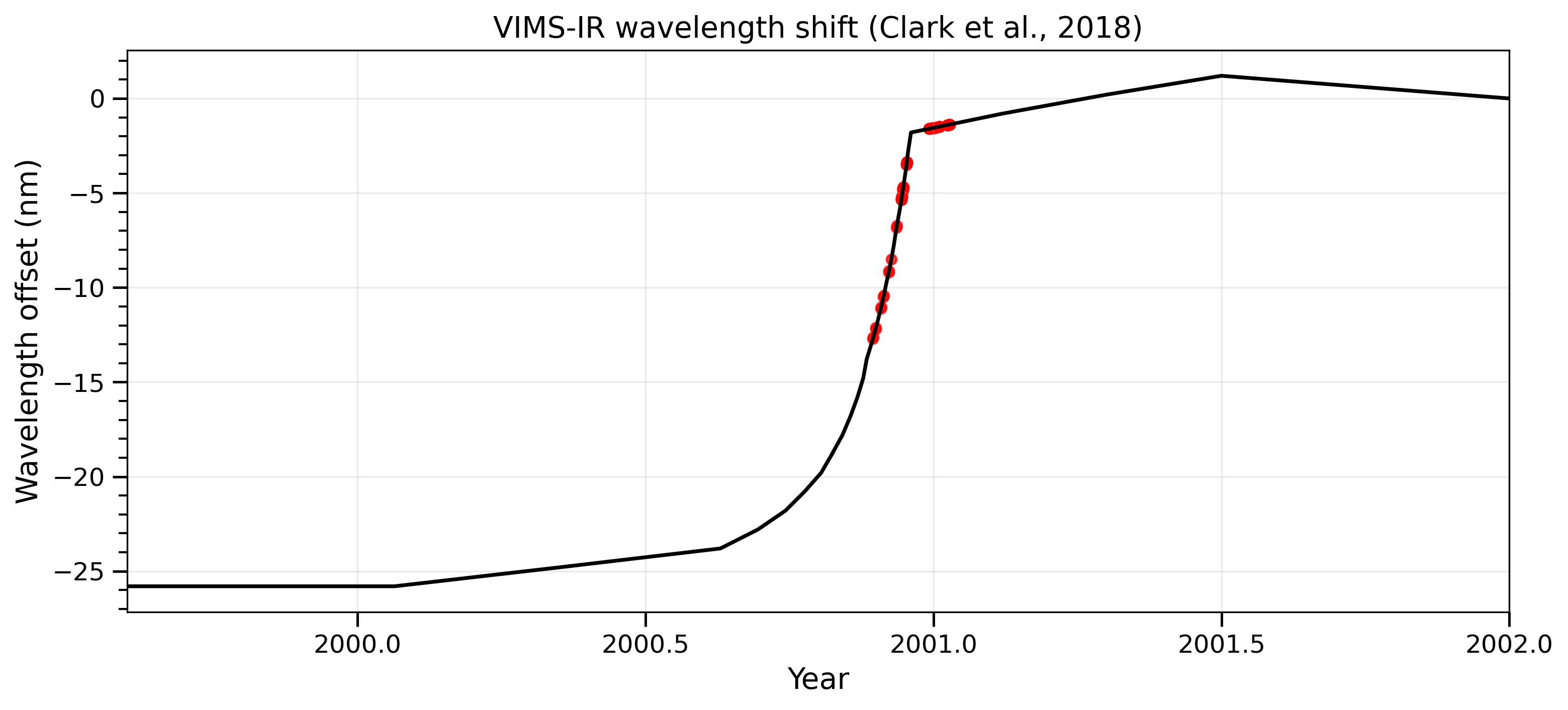}
\caption{Time-dependent VIMS-IR wavelength shift recommended by \citet{Clark2018} in the final VIMS wavelength and radiometric calibration report (RC19). Red dots show the corresponding shift values for the IR cubes in our data set, obtained by interpolation to each cube's observation time.}
\label{fig:ir_wavelength_shift}
\end{figure}

For the IR channel, \citet{Clark2018} provide a time-dependent wavelength offset referenced to the standard 2004.0 wavelength calibration, implemented as a uniform additive shift to the IR wavelength grid. The resulting time-dependent wavelength shift is shown in Figure~\ref{fig:ir_wavelength_shift}. However, \citet{Sromovsky2010} investigated the possibility of wavelength errors by repeating their spectral fits using shifted wavelength scales and evaluating the goodness of fit as a function of the applied shift. For the Jupiter flyby observations, they found that any wavelength adjustment required was within 1~nm and concluded that no further revision of the wavelength scale was necessary for their analysis.

This level of shift is smaller than the offset implied by the \citet{Clark2018} time-dependent correction near the end of year 2000 ($-1.8$~nm), and much smaller than the larger shifts suggested by \citet{McCord2004}, who reported VIMS wavelength shifts of roughly 12--22~nm based on comparisons with Galileo/NIMS. This difference may partly reflect the difficulty of inferring small wavelength offsets from inter-instrument comparisons when the data sets differ in spectral sampling, line-spread function, and absolute calibration. Notably, the VIMS-IR spectral resolution itself is of order 11--22~nm, comparable to the magnitude of the shifts under discussion. For this reason, approaches based on internal fits to the same data set while scanning an imposed wavelength shift, as in \citet{Sromovsky2010}, may provide a more robust estimate for the observations considered here.

In many practical applications, the nominal VIMS-IR wavelength solution is sufficient, and the need for a correction can be analysis-dependent. We therefore do not apply any wavelength shift to the spectra in this data set by default. Instead, we include in the FITS header of each IR cube the wavelength offset recommended by \citet{Clark2018}, interpolated to the cube observation time, leaving end users free to decide whether to apply the correction for their specific science case.

It should also be noted that the order-sorting filter joints near 1.6, 3.0, and 3.9~$\mu$m are fixed with respect to the detector geometry and therefore do not shift with time, as indicated by \citet{Clark2018}. The FWHM values are taken directly from the RC19 standard set and are not modified by the wavelength-shift correction described above.

Finally, certain IR spectral regions are known to be less reliable because of instrumental artifacts and calibration limitations. \citet{Sromovsky2010} highlight (i) problematic responsivity corrections near order-sorting filter joints, especially around 1.64~$\mu$m, with significant errors over roughly 1.60--1.68~$\mu$m, and near the joint at $\sim$2.98~$\mu$m; and (ii) an additional artifact affecting the $\sim$2.3~$\mu$m region, likely related to stray light in the spectrometer, with associated discrepancies also noted near $\sim$2.58~$\mu$m in comparisons with independent spectra. These wavelength intervals should therefore be treated with caution in quantitative interpretation. In addition, the primary FITS header records band-level quality-control flags inherited from the RC19 calibration compilation of \citet{Clark2018} (e.g., \texttt{ORSORT}, \texttt{HOTPIX}, \texttt{NOISY}), which list spectral channels potentially affected by specific instrumental issues such as order-sorting filter junctions, hot pixels, or elevated noise. These keywords may be used to identify and, if necessary, mask or exclude the corresponding bands in downstream analyses.

\section{Calibration validation}
\label{sec:calibration_validation}

\subsection{VIS Channel}

\begin{figure}[ht!]
\centering
\includegraphics[width=0.65\columnwidth]{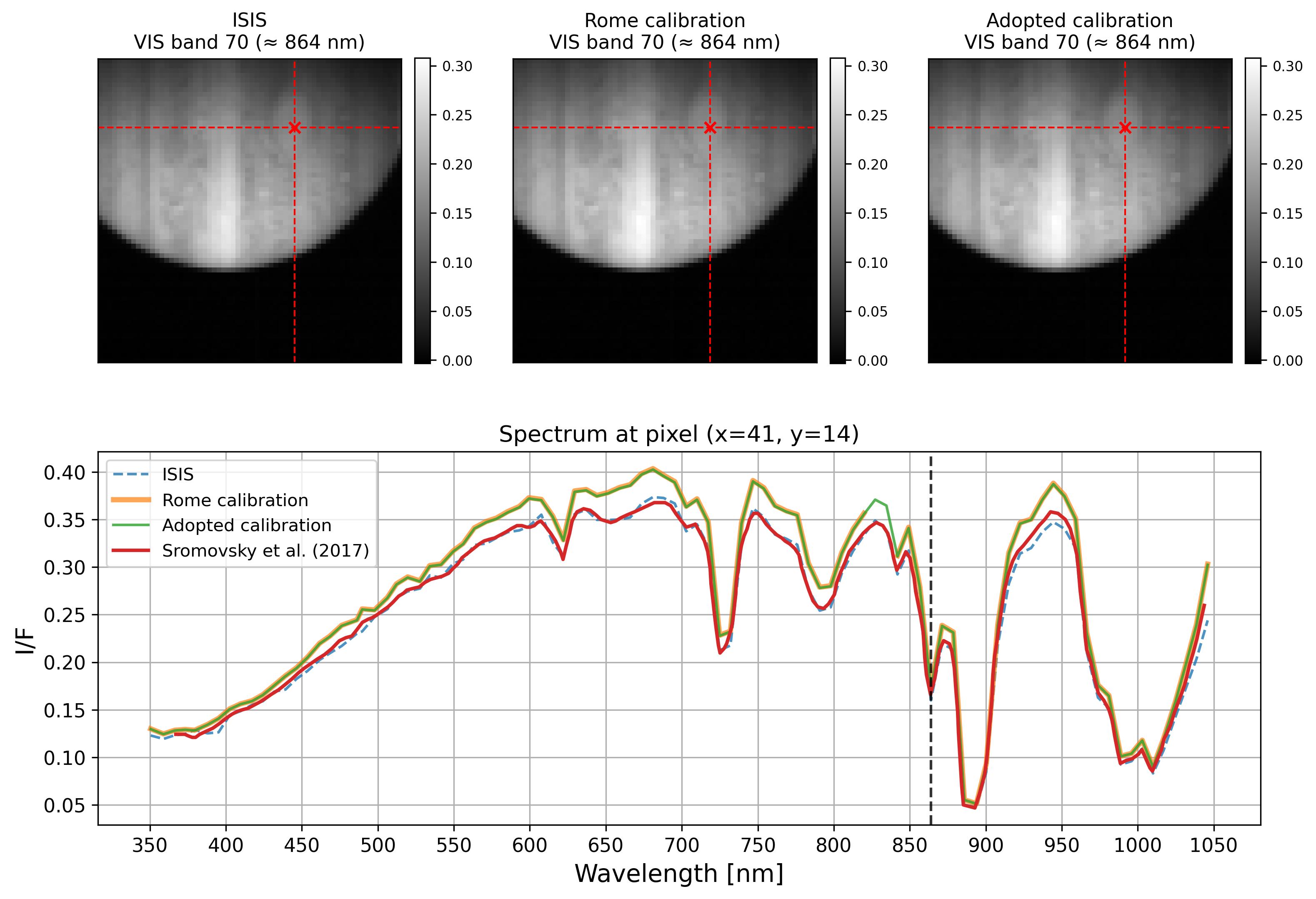}
\caption{Comparison of VIS spectra for the Great Red Spot in cube \texttt{C1356976257\_3\_vis.fits}. The three upper panels show the same VIS image at band 70 ($\lambda = 864$ nm) for the ISIS calibration, the Rome calibration, and the adopted calibration, respectively. The red dashed lines mark the pixel used for the spectral comparison. The lower panel shows the spectrum extracted at that pixel ($x = 41$, $y = 14$) for the three calibrations, together with the reference spectrum of \citet{Sromovsky2017}. The black dashed vertical line indicates the band used for the image display.}
\label{fig:vis_grs_comparison}
\end{figure}

To validate the calibration adopted in this work, in Figure~\ref{fig:vis_grs_comparison} we compare spectra of the Great Red Spot (GRS) extracted from cube \texttt{C1356976257\_3\_vis.fits} using three calibrations: (i) ISIS, (ii) the Rome-calibrated product provided by INAF--IAPS, and (iii) the calibration adopted in this work. These spectra are compared with the reference GRS spectrum published by \citet{Sromovsky2017}, all sampled at the same pixel. The comparison shows that the adopted calibration reproduces very accurately both the spectral shape and the absolute $I/F$ level of the Rome calibration, indicating that the responsivity-matrix implementation in our pipeline is correct. The spectrum from \citet{Sromovsky2017} agrees reasonably well with the ISIS curve, indicating that these two references are mutually consistent, but they both lie slightly below the adopted calibration in the visible range.

\begin{figure}[h]
\centering
\includegraphics[width=0.55\columnwidth]{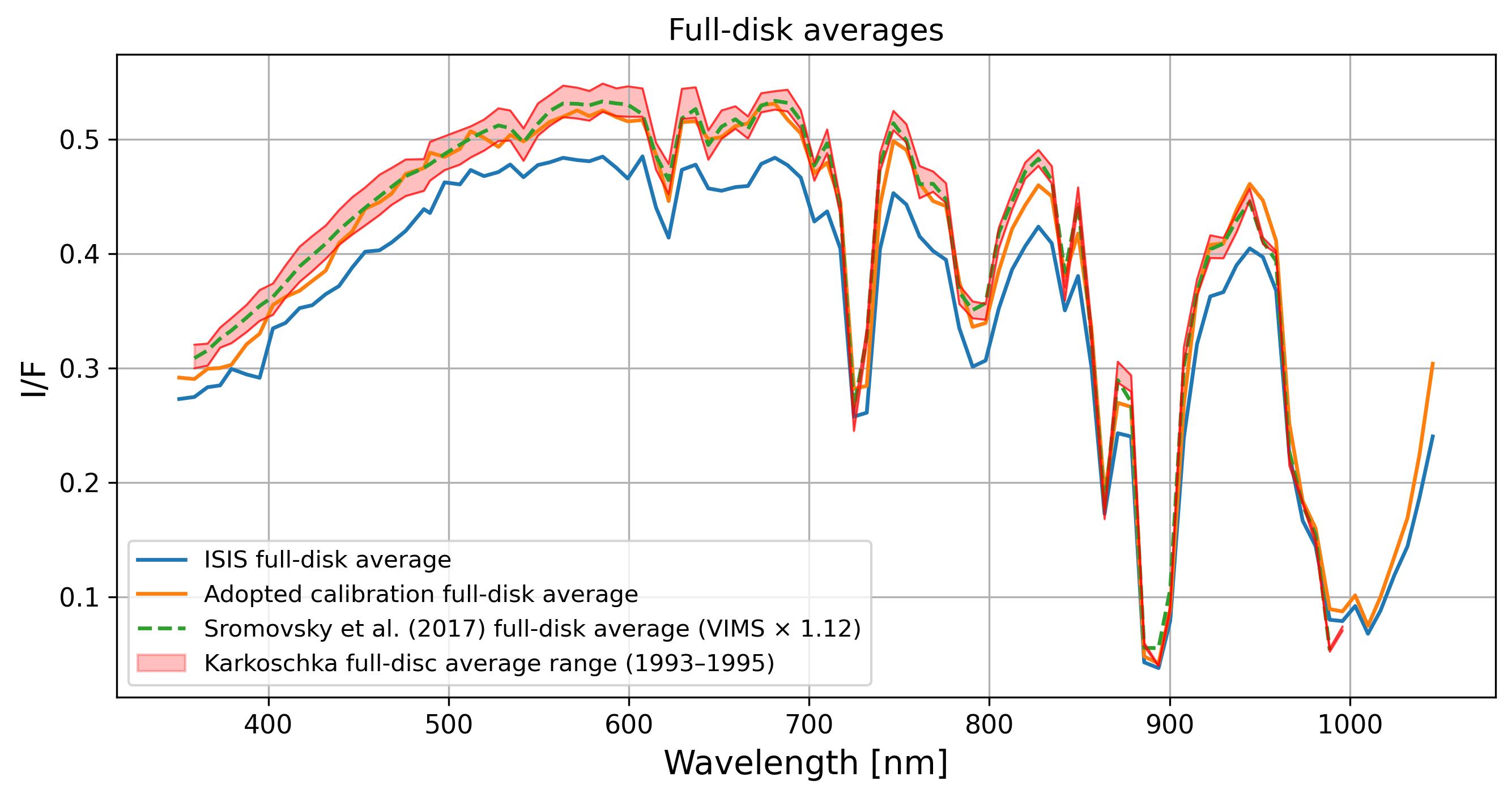}
\caption{Full-disk average comparison for the VIMS-VIS cube \texttt{C1354610545\_2\_vis.fits}. The blue curve shows the full-disk average obtained from the standard ISIS calibration, and the orange curve shows the corresponding result from the adopted calibration. The green curve is the scaled ``VIMS $\times$ 1.12'' full-disk reference spectrum from \citet{Sromovsky2017}. The red shaded region indicates the envelope spanned by the Jupiter full-disk spectra of \citet{Karkoschka1994} and \citet{Karkoschka1998}.}
\label{fig:vis_fulldisc_comparison}
\end{figure}

To further assess the radiometric consistency of the adopted calibration, we compared full-disk averages derived from the ISIS standard pipeline and from our calibrated products against the disk-integrated spectra reported by \citet{Sromovsky2017} and by \citet{Karkoschka1994,Karkoschka1998} (Figure~\ref{fig:vis_fulldisc_comparison}). \citet{Sromovsky2017} argued that the VIMS radiometric calibration appears to require an increase of approximately 10\%, and therefore presented their full-disk spectrum as VIMS $I/F$ multiplied by a factor of 1.12. In our comparison we use this already scaled curve (labelled ``VIMS $\times$ 1.12'') as the Sromovsky reference. The Karkoschka full-disk averages for Jupiter at the 1993 and 1995 epochs are also shown. We represent the range spanned by these two data sets as a shaded envelope, which can be regarded as an empirical constraint on the absolute level and spectral shape of Jupiter's visible full-disk albedo. The comparison shows that the standard ISIS calibration lies systematically below the Karkoschka envelope over much of the VIMS range, in line with the conclusion by \citet{Sromovsky2017} that an upward rescaling of the original VIMS calibration is required. By contrast, the adopted calibration developed in this work tracks both the Sromovsky (VIMS $\times$ 1.12) curve and the Karkoschka range very closely, without the need for any additional scaling. This indicates that the revised calibration effectively incorporates the $\sim$10\% correction suggested by \citet{Sromovsky2017}.

As a methodological note, the spectrum and full-disk average from \citet{Sromovsky2017} were not taken from an original digital data file, but were instead reconstructed by manually sampling the curves shown in their published figures. Consequently, these curves should not be regarded as exact reproductions of the original spectrum and full-disk average.

\subsection{IR Channel}
\label{sec:validation_IRchannel}

Some IR-channel cubes acquired during the Jupiter approach phase show clearly over-scaled spectra when calibrated with ISIS, with $I/F$ values exceeding unity in the solar-reflected part of the spectrum. We traced this behavior to the $I/F$ normalization implemented in the ISIS VIMS calibration application \texttt{vimscal}. To convert to units of $I/F$, \texttt{vimscal} estimates the Sun--target distance term ($d^2$) by sampling a fixed set of reference pixels, namely the cube center, the four corners, and the midpoints of the four edges. For approach observations, the Jovian disk can occupy only a small, off-centered fraction of the frame, so these reference pixels may all fall outside the planet and therefore have no valid surface intersection. In that situation, \texttt{vimscal} falls back to a hard-coded default $d^2$ value ($\sim$81.6~AU$^2$, i.e., the Sun--Saturn distance squared in early 2004), which introduces an artificial multiplicative scaling of order $\sim$81.6/$d_{\rm Jup}^2 \approx 3.2$ in the resulting $I/F$.

\begin{figure}[h]
\centering
\includegraphics[width=0.85\textwidth]{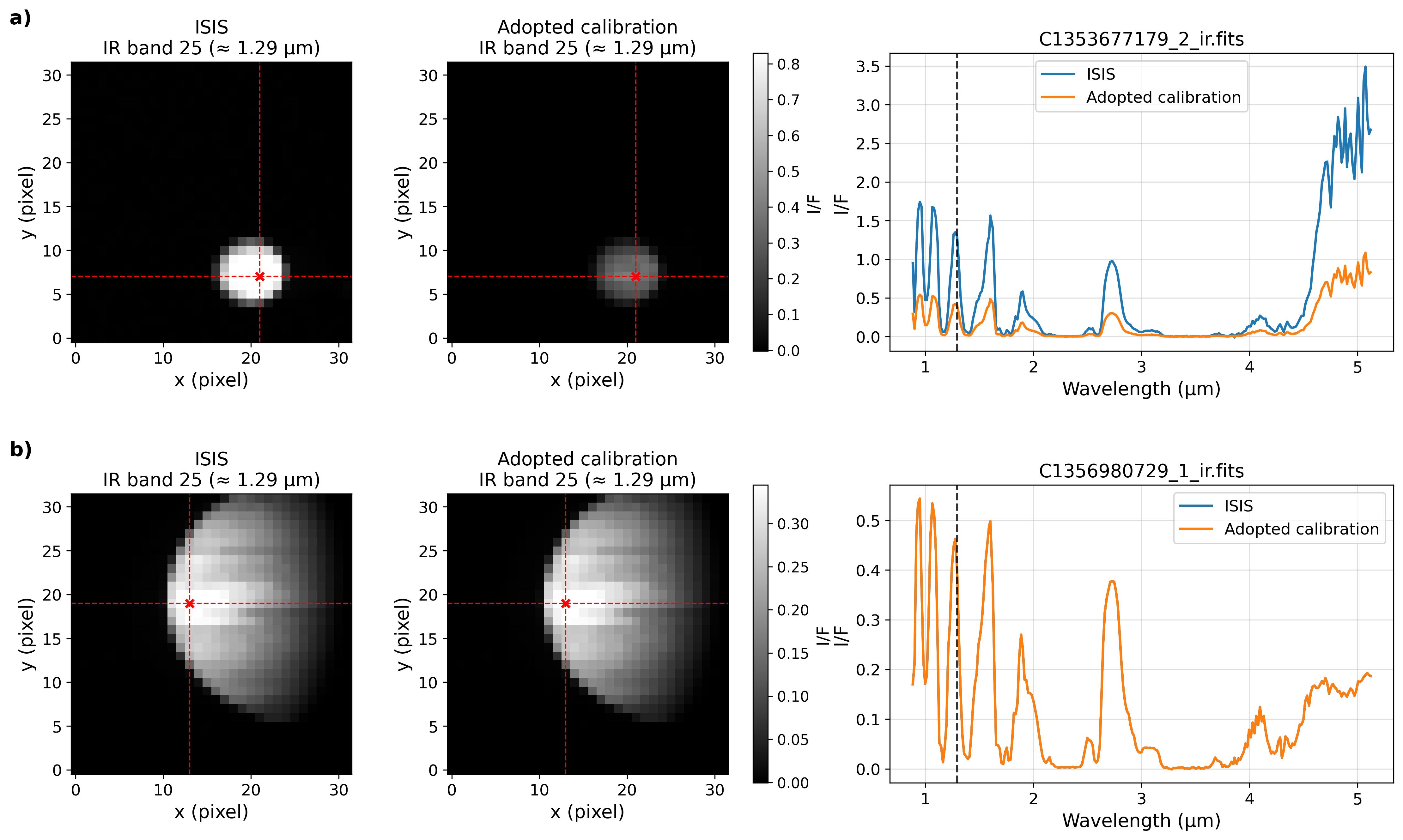}
\caption{Comparison of the ISIS standard calibration and the adopted calibration for VIMS-IR cubes acquired during (a) the Jupiter approach phase and (b) the flyby phase. Panel (b) shows a case unaffected by the Sun--target distance issue, for which agreement is expected because both reductions are based on the standard VIMS/RC19 calibration files.} The black dashed vertical line indicates the band used for the image display.
\label{fig:ir_isis_comparison}
\end{figure}

The calibration adopted in this work, based on the custom responsivity matrices described in Section~\ref{sec:ir_general_pipeline}, avoids this failure mode by computing Jupiter's heliocentric distance directly from the observation start time and ephemeris. This distance agrees with the on-disk ISIS geometry to within 0.001\%, which is more than sufficient for the radiometric precision required here. The issue was identified only after calibrating the same cubes with our custom pipeline and comparing the results, which revealed a nearly constant multiplicative offset of $\sim$3.2 for the affected observations. Panel~(a) of Figure~\ref{fig:ir_isis_comparison} shows an over-scaled spectrum obtained with ISIS for an approach-phase cube, together with the satisfactory output from our pipeline. The problem is not seen in cubes acquired during the flyby phase, where the Jovian disk typically spans the central regions of the frame and the adopted calibration closely matches the ISIS results, as shown in panel~(b) of Figure~\ref{fig:ir_isis_comparison}.

\begin{figure}[h]
\centering
\includegraphics[width=0.6\columnwidth]{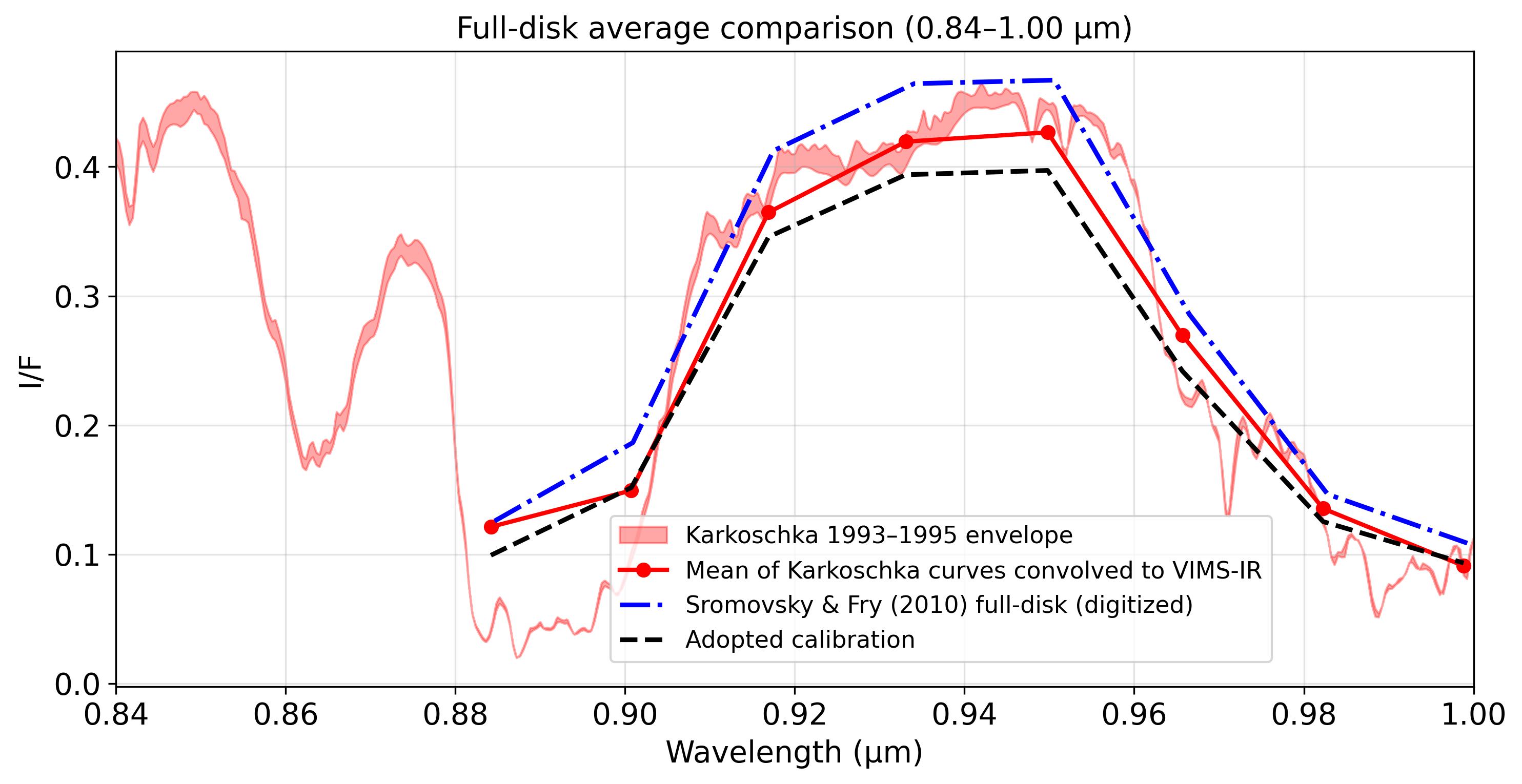}
\caption{Comparison of the adopted calibration for cube \texttt{C1355256529\_3\_ir.fits} with the full-disk averages of \citet{Karkoschka1994}, \citet{Karkoschka1998}, and \citet{Sromovsky2010}. The red curve shows the mean of the \citet{Karkoschka1994} and \citet{Karkoschka1998} spectra after convolution to the VIMS-IR spectral resolution.}
\label{fig:ir_fulldisc_comparison}
\end{figure}

We next compare the adopted calibration against published full-disk reference spectra (Figure~\ref{fig:ir_fulldisc_comparison}). Our VIMS full-disk curve is generally slightly lower than the \cite{Karkoschka1994,Karkoschka1998} envelope, but remains in good agreement with the Karkoschka reference spectra after convolution to the VIMS-IR spectral resolution, indicating that the wavelength-dependent behavior is consistent. The full-disk average was computed using a disk mask derived from the geometric backplanes (see Section~\ref{sec:geom_backplanes}). We note that the comparison of Figure~\ref{fig:ir_fulldisc_comparison} uses the same VIMS cube (C1355256529\_3\_ir.fits) as \citet{Sromovsky2010}. The exact mask definition affects the resulting full-disk spectrum, and \citet{Sromovsky2010} do not describe the criterion used to define their disk mask. Therefore, part of the difference between the two curves may reflect differences in the full-disk averaging procedure, in addition to possible differences between ISIS calibration versions. The comparatively weaker VIMS absorption in the CH$_4$ band centered at 0.89~$\mu$m is mainly a sampling effect, because the band minimum is not fully captured by the IR spectral grid, whereas the VIS channel resolves this feature more completely, as shown in Figure~\ref{fig:vis_fulldisc_comparison}.

\begin{figure}[ht!]
\centering
\includegraphics[width=0.65\columnwidth]{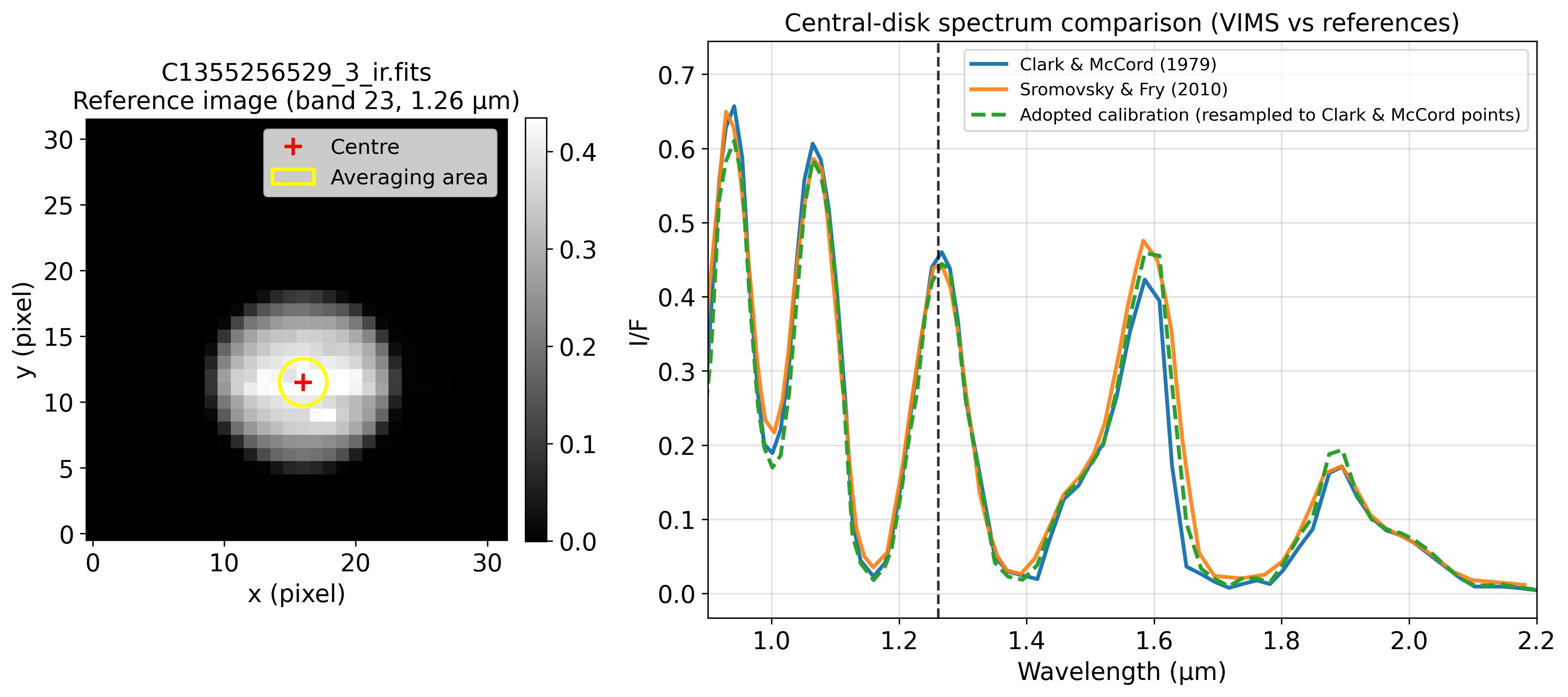}
\caption{Comparison of the adopted VIMS-IR calibration with the central-disk averages of \citet{Clark1979} and \citet{Sromovsky2010}. The left panel shows the averaging region. Spectra have been resampled to the spectral resolution of \citet{Clark1979}.}
\label{fig:ir_centraldisk_comparison}
\end{figure}

We also compare the adopted calibration with the central-disk spectra of \citet{Clark1979} and \citet{Sromovsky2010} in Figure~\ref{fig:ir_centraldisk_comparison}, finding overall good agreement. A localized mismatch around 1.6~$\mu$m appears to be related to the junction between the VIMS order-sorting filters in the detector assembly. The disk area used for our central average is smaller than that used by \citet{Clark1979}, as we restrict the region to avoid including the Great Red Spot, which is not present in the \citet{Clark1979} image. For consistency, we computed our central-disk average from the same VIMS observation used by \citet{Sromovsky2010} (\texttt{C1355256529\_3\_ir.fits}). The averaging region used by \citet{Sromovsky2010} is not specified. It should also be noted that both the \citet{Clark1979} and \citet{Sromovsky2010} reference curves were reconstructed by manually sampling their published figures, introducing an additional source of uncertainty.

\begin{figure}[ht!]
\centering
\includegraphics[width=0.5\columnwidth]{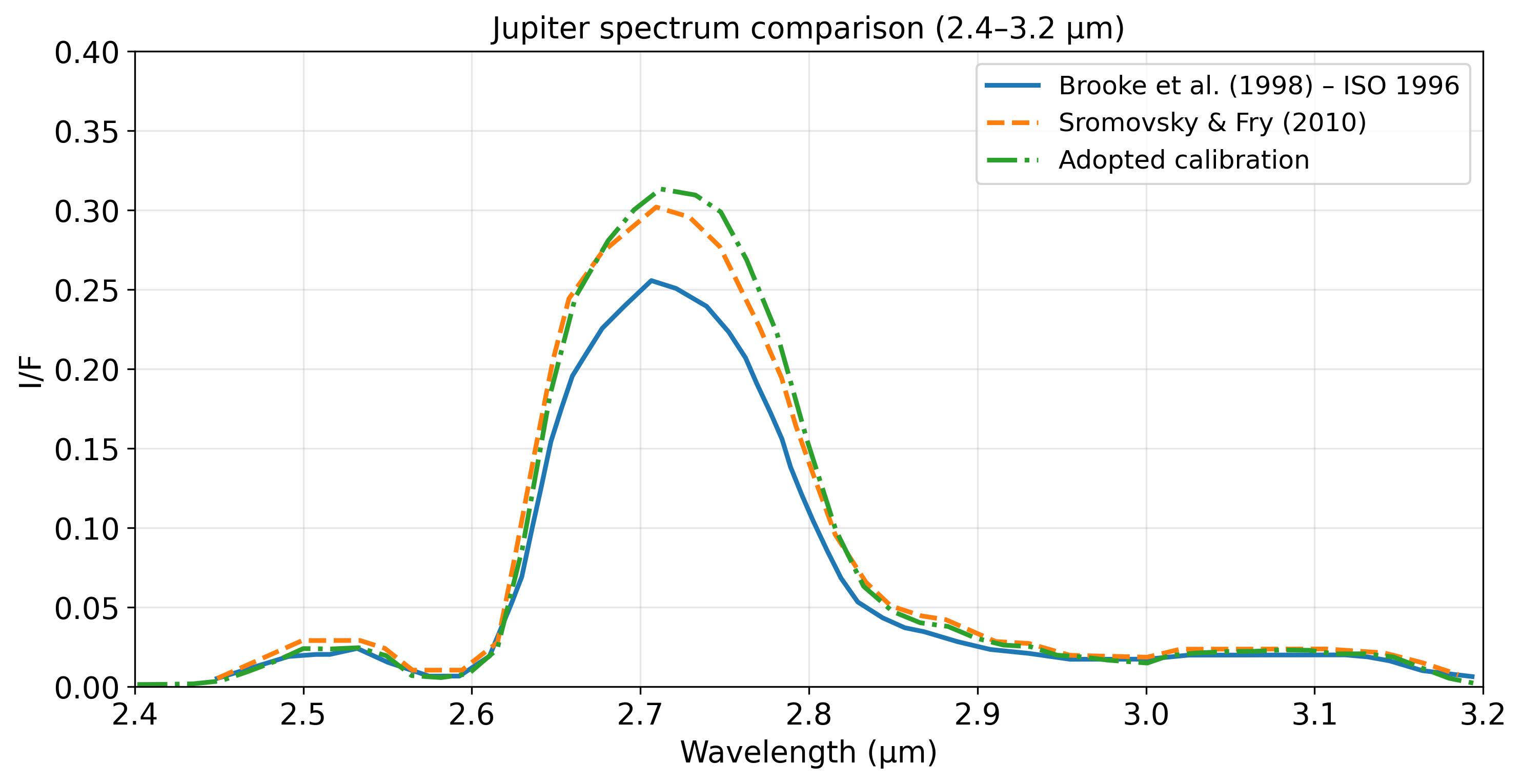}
\caption{Jupiter spectrum comparison in the 2.4--3.2~$\mu$m range. The VIMS-IR spectrum from this work, using the adopted calibration and a Brooke-like field of view, is compared with the ISO spectrum of \citet{Brooke1998} and the VIMS-IR reference spectrum of \citet{Sromovsky2010}.}
\label{fig:ir_brooke_comparison}
\end{figure}

In Figure~\ref{fig:ir_brooke_comparison}, we compare our VIMS spectrum in the 2.4--3.2~$\mu$m range with the reference VIMS spectrum of \citet{Sromovsky2010} and the ISO spectrum of \citet{Brooke1998}. As noted by \citet{Sromovsky2010}, the differences relative to ISO can plausibly be attributed to the different observing geometries (phase angle $\approx$2.5$^\circ$ for VIMS versus $\approx$11$^\circ$ for ISO), to temporal variability in Jupiter's clouds and hazes, and to the absolute calibration uncertainty of ISO, which can be as large as $\sim$12\%.

\begin{figure}[h]
\centering
\includegraphics[width=0.5\columnwidth]{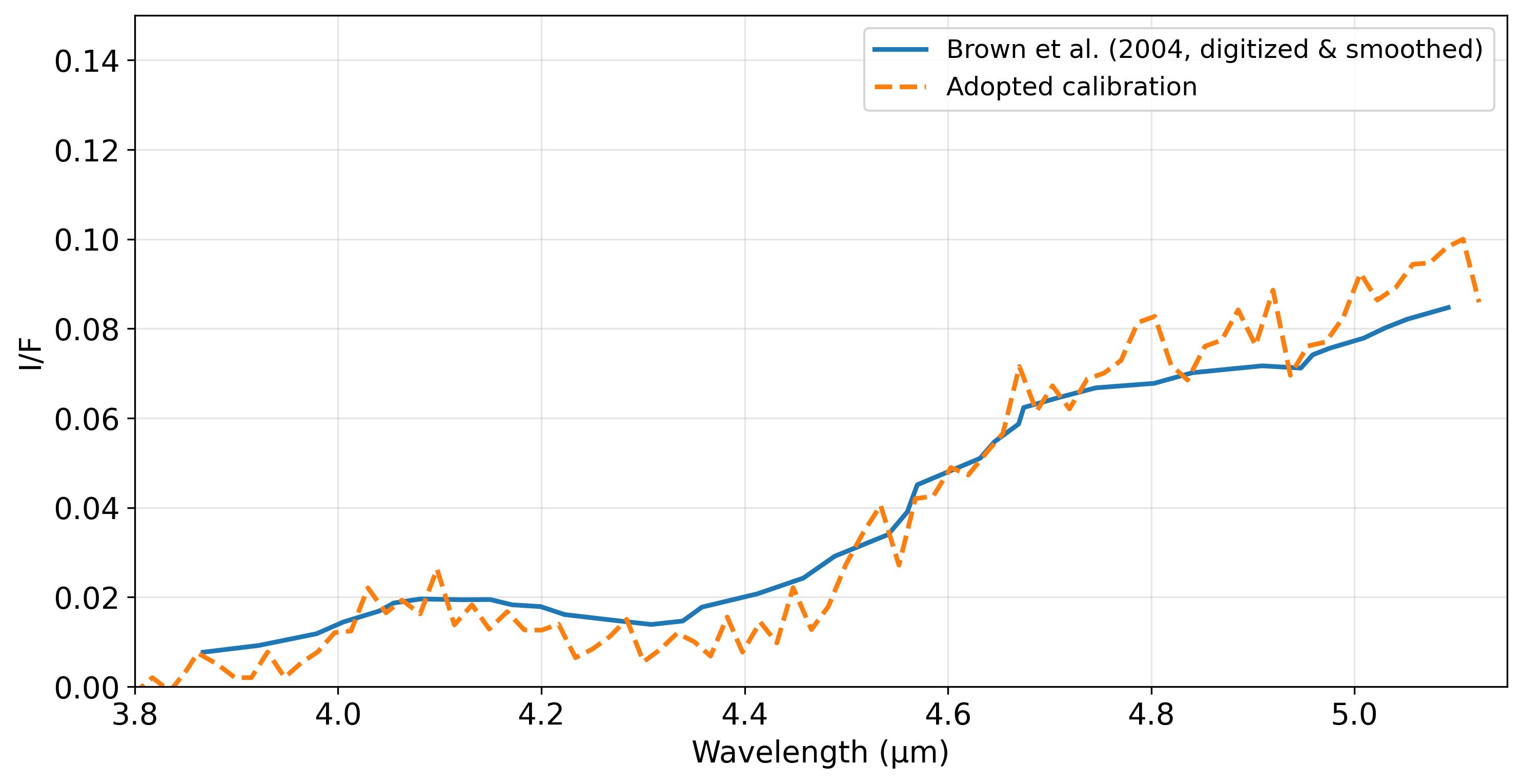}
\caption{Thermal-regime comparison of a VIMS-IR pixel spectrum extracted over the Great Red Spot from cube \texttt{C1356907330\_3\_ir.fits} with the Great Red Spot spectrum of \citet{Brown2004}. The \citet{Brown2004} curve was digitized from the published figure and then smoothed to suppress digitization-related fluctuations.}
\label{fig:ir_thermal_comparison}
\end{figure}

Finally, we analyze the thermal part of the VIMS-IR spectrum in Figure~\ref{fig:ir_thermal_comparison}. A direct comparison in this wavelength range is difficult because the thermal emission shows strong spatial and temporal variability across Jupiter, so spectra depend sensitively on the region sampled. For this reason, we compare our data with a VIMS-IR spectrum of the Great Red Spot reported by \citet{Brown2004}. Since this reference spectrum was reconstructed by manually digitizing a published figure over a relatively narrow wavelength interval, the sampled curve is subject to additional uncertainty, including imperfect capture of the small-scale peaks and troughs and small errors in the wavelength axis. We therefore smoothed the digitized curve in order to suppress these digitization-related fluctuations and retain only its broad spectral behavior. The observation corresponding to that spectrum was not specified, so we inspected all available VIMS-IR cubes to identify a thermal-regime spectrum that best matches the \citet{Brown2004} reference. The closest agreement is obtained for cube \texttt{C1356907330\_3\_ir.fits}, in which the Great Red Spot lies close to the terminator. Overall, the agreement is reasonably good.

\section{Geometry Backplanes}
\label{sec:geom_backplanes}

Geometry backplanes provide, for each pixel in a VIMS cube, the local observing and illumination geometry. They are essential for comparing different cubes, correcting for viewing conditions, and interpreting spatial and spectral variability in a physically meaningful way. In this work, the geometry backplanes were generated from SPICE kernels using the ISIS application \texttt{phocube}. The final calibrated FITS products include backplanes for the phase angle, emission angle, incidence angle, planetocentric latitude, positive-east longitude, pixel resolution, and azimuth angle. The angular backplanes are given in degrees, whereas pixel resolution is given in meters.

\begin{figure}[ht!]
\centering
\includegraphics[width=0.45\columnwidth]{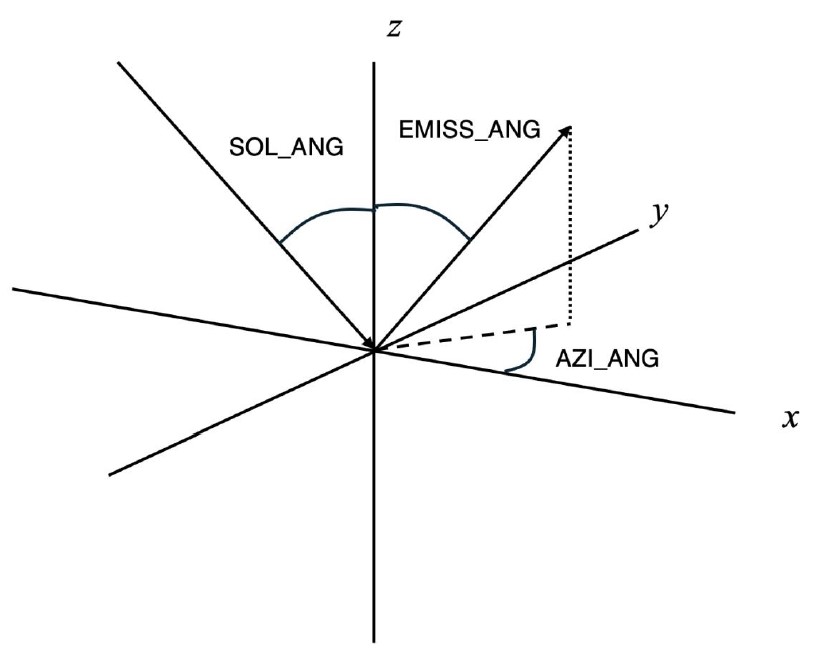}
\caption{Definition of the incidence angle $\theta_0$, emission angle $\theta$, phase angle $\alpha$, and azimuth angle $\phi$ used to compute the azimuth backplane following the adopted convention.}
\label{fig:angle_definitions}
\end{figure}

The azimuth angle, as defined below, is not provided directly by ISIS and was therefore computed separately. We adopt the convention used by the NEMESIS radiative transfer suite \citep{Irwin2008}, as illustrated in Figure~\ref{fig:angle_definitions}. Let the solar incidence angle be $\theta_0$, the emission angle be $\theta$, the azimuth angle be $\phi$, and the phase angle be $\alpha$. In a local Cartesian basis $(\mathbf{i},\mathbf{j},\mathbf{k})$, the unit vectors toward the Sun and toward the observer may be written as

\begin{equation}
\mathbf{v}_1 = (-\sin\theta_0,\,0,\,\cos\theta_0),
\end{equation}

\begin{equation}
\mathbf{v}_2 = (\sin\theta\cos\phi,\,\sin\theta\sin\phi,\,\cos\theta).
\end{equation}

The phase angle follows from the dot product of these two vectors,

\begin{equation}
\cos\alpha = \mathbf{v}_1 \cdot \mathbf{v}_2
= -\sin\theta_0 \sin\theta \cos\phi + \cos\theta_0 \cos\theta,
\end{equation}

which can be rearranged to give

\begin{equation}
\cos\phi =
\frac{\cos\theta_0 \cos\theta - \cos\alpha}
{\sin\theta_0 \sin\theta}.
\end{equation}

\subsection{VIS Channel}

\subsubsection{Detection of Misalignment Between Radiometric and Geometric Disks}

For a significant fraction of VIS cubes, a systematic offset was found between the disk defined by the VIS reflectivity image (the ``radiometric disk'') and the disk defined by the geometric backplanes (the ``geometric disk''). Ideally, these two disks should coincide. To detect and quantify the misalignment, we implemented an automatic procedure based on binary masks derived from both radiometric and geometric information.

A VIS band with high signal-to-noise ratio was selected from the calibrated cube, near 680~nm, where Jupiter is particularly bright. An $I/F$ threshold of 0.05 was applied to construct a binary radiometric mask, in which pixels belonging to the planetary disk were assigned a value of 1 and background pixels a value of 0. In parallel, a geometric mask was built from the emission-angle backplane by selecting pixels with valid emission-angle values.

The areas of the two masks were then compared. We define the relative area difference as

\begin{equation}
\Delta A_{\rm rel} =
\frac{|A_{\rm rad} - A_{\rm geo}|}{\max(A_{\rm rad},A_{\rm geo})},
\end{equation}

where $A_{\rm rad}$ and $A_{\rm geo}$ are the numbers of pixels equal to 1 in the radiometric and geometric masks, respectively. This quantity was used as an indicator of whether the two masks represented similar, nearly complete disks, or whether at least one of them was strongly truncated at the edge of the field of view.

In practice, values of $\Delta A_{\rm rel} < 0.14$ were taken to indicate similar mask areas, corresponding to complete or nearly complete disks, whereas values $\geq 0.14$ were interpreted as markedly different mask areas, typically associated with strongly truncated partial disks. This threshold was determined empirically by testing a range of candidate values. For each candidate threshold, cubes were assigned to one of the two overlap metrics described below, and the resulting offsets were inspected for representative cases. The adopted value, 0.14, was the one for which this assignment most consistently produced correct disk alignments.

Depending on the value of $\Delta A_{\rm rel}$, two different overlap metrics were used. When $\Delta A_{\rm rel} < 0.14$, we used the simple intersection between the two masks, i.e., the number of pixels equal to 1 in both masks. When $\Delta A_{\rm rel} \geq 0.14$, the simple intersection becomes less reliable because one mask may be largely contained within the other even when the alignment is poor. In such cases we used instead the intersection-over-union (IoU) metric,

\begin{equation}
{\rm IoU} = \frac{A_{\rm int}}{A_{\rm union}},
\end{equation}

where $A_{\rm int}$ is the number of pixels common to both masks and $A_{\rm union}$ is the number of pixels belonging to either mask.

The radiometric mask was shifted over a grid of integer offsets $(\Delta x,\Delta y)$ within a limited search window around zero, typically $\pm 10$ pixels in each direction. For each candidate offset, either the intersection or the IoU was evaluated, depending on the mask-area regime. The offset that maximized the selected metric was adopted as the best estimate of the true misalignment between the radiometric and geometric disks.

\subsubsection{Pointing Correction With \texttt{deltack}}

The measured offset was not applied as a simple post-processing shift of the images or backplanes. Instead, it was used to refine the camera pointing within ISIS, so that the geometry became intrinsically consistent with the VIS cube and missing geometry could be recovered.

A robust centroid of the Jovian disk was computed from the radiometric mask and rounded to integer pixel coordinates $(\mathrm{line}_1,\mathrm{samp}_1)$. Using the original camera model and the measured offset $(\Delta x,\Delta y)$, we derived the planetary coordinates $(\mathrm{lat}_1,\mathrm{lon}_1)$ that this reference pixel should have in the corrected geometry. The ISIS application \texttt{deltack} was then used in direct mode to update the camera pointing such that the image point $(\mathrm{line}_1,\mathrm{samp}_1)$ mapped exactly to $(\mathrm{lat}_1,\mathrm{lon}_1)$. After this adjustment, SPICE-based geometry calculations, including recomputation of \texttt{phocube}, produced new backplanes intrinsically aligned with the VIS spectral cube.

\begin{figure}[ht!]
\centering
\includegraphics[width=0.6\columnwidth]{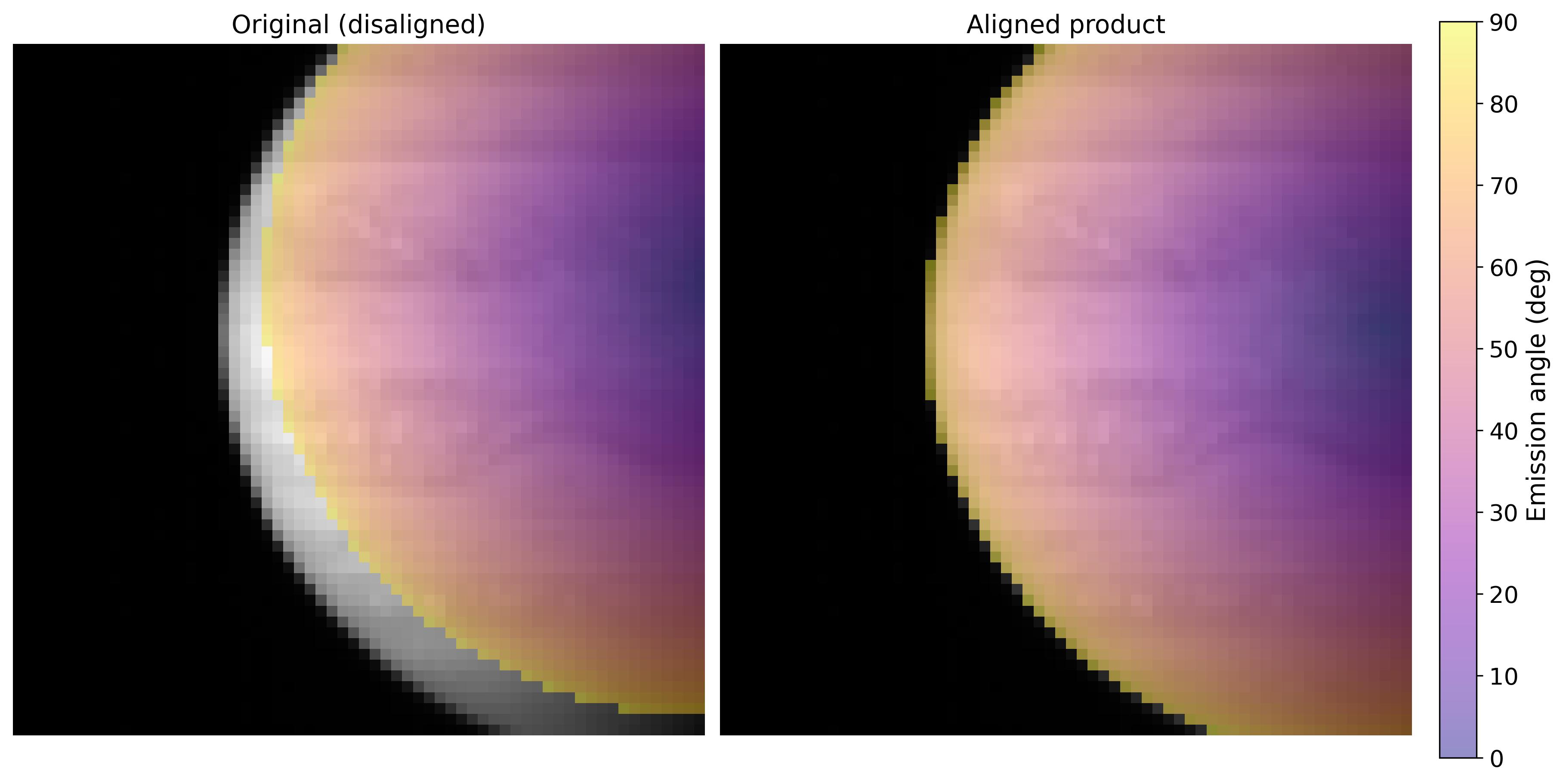}
\caption{Example VIS cube before and after automatic alignment of the geometric backplanes with respect to the spectral cube. The emission angle is overlaid on the gray-scale $I/F$ image VIS band 46 (680~nm).}
\label{fig:vis_alignment_example}
\end{figure}

This approach yields a self-consistent geometric solution at the level of the camera model rather than at the level of individual images or backplanes.  An example of the alignment before and after the pointing correction is shown in Figure~\ref{fig:vis_alignment_example}.

\subsubsection{Residual Uncertainty}

Although the automatic method successfully corrected most VIS cubes, some cases, approximately 20\%, required manual adjustment. In these cases, all acceptable offsets were inspected visually and the preferred solution was selected from the limb alignment.

Even after these refinements, the final alignment is not perfect in all cases. In particular, the effective size of the radiometric disk is not always identical to that of the geometric disk, and in low-resolution cubes a good match on one part of the limb can still leave a residual mismatch of about one pixel on the opposite side. We therefore adopt a characteristic residual geometric uncertainty of approximately one pixel for the VIS channel.

\subsection{IR Channel}

For the IR channel, the same overlap analysis between the radiometric and geometric disks was applied as for the VIS channel. In all cases, the best solution corresponded to a zero offset, indicating that the original pointing of the IR cubes was already correct and that no systematic pixel shift between the IR images and the IR geometric backplanes was present.

Nevertheless, a residual pointing uncertainty of about one pixel remains in the IR channel. As in the VIS channel, the effective size of the radiometric disk is not exactly the same as that of the geometric disk. This behavior appears to be intrinsic to the data and/or to the ISIS camera model rather than the result of an uncorrected pointing offset. We therefore adopt the same characteristic geometric uncertainty, of order one pixel, for the IR channel.

\section{Conclusions}

We have presented a processing workflow for Cassini VIMS Jupiter cubes, together with the resulting catalog of calibrated VIS and IR products. Starting from the raw archive data, the workflow produces multi-extension FITS files that combine radiometrically calibrated spectral cubes, wavelength information, and geometry backplanes in a uniform format suitable for subsequent scientific analysis. In the VIS channel, the adopted calibration effectively incorporates the correction reported by \citet{Sromovsky2017}. In the IR channel, the workflow resolves a subset of problematic cases not satisfactorily handled by the default ISIS pipeline. Overall, the resulting products are broadly consistent with independent reference spectra from the literature. The final data set, together with the responsivity matrices used in the calibration, is publicly available \citep{AnguianoArteaga2026Zenodo} and provides a validated and homogeneous basis for future studies of Jupiter using Cassini VIMS observations.

\begin{acknowledgments}
We are very grateful to G.~Filacchione for constructing and sharing the responsivity matrices used in the VIS-channel calibration.

This work was supported by grant PID2023--149055NB--C31, funded by MICIU/AEI/10.13039/501100011033 and by FEDER, UE, and also by Elkartek KK-2025/00106.

A.~Anguiano-Arteaga was supported by the \textit{Programa de Perfeccionamiento de Personal Investigador Doctor 2024--2027} of the Basque Government.
\end{acknowledgments}

\begin{contribution}
A.~Anguiano-Arteaga carried out the data processing and calibration work, generated the final data products, performed the validation analyses, and wrote the manuscript. P.~Irwin and S.~P\'erez-Hoyos conceived the project and contributed to the interpretation of the results and to the development and revision of the manuscript. D.~Grassi and E.~D'Aversa provided guidance on the VIS-channel calibration, particularly in connection with the calibration approach developed at INAF--IAPS. All authors reviewed the draft and contributed to the final version of the manuscript.
\end{contribution}

\software{
ISIS \citep{Anderson2004},
Astropy \citep{Astropy},
SAOImage DS9 \citep{Joye2003},
QFitsView \citep{Ott2012}
}

\appendix

\section{Summary of final products}
\label{appendix:summary_products}

Table~\ref{tab:final_products} summarizes the final calibrated Cassini VIMS Jupiter products delivered in this work. The final calibrated products follow the naming pattern \texttt{C\#\#\#\#\#\#\#\#\#\_\#\_[vis/ir].fits}. The numeric block is inherited from the original VIMS filename and corresponds to the spacecraft clock time tag of the observation. The suffix after the underscore is likewise inherited from the original VIMS naming convention and denotes the version number of the archive product. The initial \texttt{C} is added here to distinguish the calibrated FITS products from the original VIMS archive files. This naming scheme follows the original VIMS archive convention described in the VIMS Archive Volume SIS \citep{Brown2005}.

\newpage
\startlongtable
\begin{deluxetable*}{lccccc}
\tabletypesize{\scriptsize}
\tablewidth{0pt}
\tablecaption{Summary of final calibrated Cassini VIMS Jupiter products\label{tab:final_products}}
\tablehead{
\colhead{File name} &
\colhead{Date} &
\colhead{Start time (UT)} &
\colhead{Exposure time (s)} &
\colhead{Sampling mode} &
\colhead{Mean phase angle ($^\circ$)}
}
\startdata
\texttt{C1353964221\_2\_vis.fits} & 2000-11-26 & 20:59:02 & 1.28 & HI-RES & 13.9 \\
\texttt{C1354179743\_2\_vis.fits} & 2000-11-29 & 08:51:03 & 1.28 & HI-RES & 12.8 \\
\texttt{C1354323264\_2\_vis.fits} & 2000-12-01 & 00:43:03 & 1.28 & HI-RES & 11.9 \\
\texttt{C1354395024\_2\_vis.fits} & 2000-12-01 & 20:39:02 & 1.28 & HI-RES & 11.4 \\
\texttt{C1354610545\_2\_vis.fits} & 2000-12-04 & 08:31:02 & 1.28 & HI-RES & 9.8 \\
\texttt{C1354754066\_2\_vis.fits} & 2000-12-06 & 00:23:02 & 1.28 & HI-RES & 8.5 \\
\texttt{C1354825827\_2\_vis.fits} & 2000-12-06 & 20:19:02 & 1.28 & HI-RES & 7.8 \\
\texttt{C1355041348\_2\_vis.fits} & 2000-12-09 & 08:11:02 & 1.28 & HI-RES & 5.4 \\
\texttt{C1355233813\_3\_vis.fits} & 2000-12-11 & 13:38:46 & 0.64 & NOMINAL & 2.9 \\
\texttt{C1355240579\_3\_vis.fits} & 2000-12-11 & 15:31:32 & 0.64 & NOMINAL & 2.7 \\
\texttt{C1355243962\_3\_vis.fits} & 2000-12-11 & 16:27:55 & 0.64 & NOMINAL & 2.6 \\
\texttt{C1355247345\_3\_vis.fits} & 2000-12-11 & 17:24:18 & 0.64 & NOMINAL & 2.6 \\
\texttt{C1355252743\_3\_vis.fits} & 2000-12-11 & 18:54:16 & 0.64 & NOMINAL & 2.6 \\
\texttt{C1355256529\_3\_vis.fits} & 2000-12-11 & 19:57:22 & 0.64 & NOMINAL & 2.5 \\
\texttt{C1355259911\_3\_vis.fits} & 2000-12-11 & 20:53:44 & 0.64 & NOMINAL & 2.5 \\
\texttt{C1355263294\_3\_vis.fits} & 2000-12-11 & 21:50:07 & 0.64 & NOMINAL & 2.4 \\
\texttt{C1355266677\_3\_vis.fits} & 2000-12-11 & 22:46:30 & 0.64 & NOMINAL & 2.3 \\
\texttt{C1355309532\_3\_vis.fits} & 2000-12-12 & 10:40:45 & 0.64 & NOMINAL & 1.7 \\
\texttt{C1355313318\_3\_vis.fits} & 2000-12-12 & 11:43:51 & 0.64 & NOMINAL & 1.7 \\
\texttt{C1355320083\_3\_vis.fits} & 2000-12-12 & 13:36:36 & 0.64 & NOMINAL & 1.5 \\
\texttt{C1355323466\_3\_vis.fits} & 2000-12-12 & 14:32:58 & 0.64 & NOMINAL & 1.4 \\
\texttt{C1355326849\_3\_vis.fits} & 2000-12-12 & 15:29:21 & 0.64 & NOMINAL & 1.4 \\
\texttt{C1355330232\_3\_vis.fits} & 2000-12-12 & 16:25:44 & 0.64 & NOMINAL & 1.3 \\
\texttt{C1355333615\_3\_vis.fits} & 2000-12-12 & 17:22:07 & 0.64 & NOMINAL & 1.3 \\
\texttt{C1355336998\_3\_vis.fits} & 2000-12-12 & 18:18:30 & 0.64 & NOMINAL & 1.2 \\
\texttt{C1355340381\_3\_vis.fits} & 2000-12-12 & 19:14:53 & 0.64 & NOMINAL & 1.2 \\
\texttt{C1355518209\_2\_vis.fits} & 2000-12-14 & 20:38:39 & 1.28 & HI-RES & 2.1 \\
\texttt{C1355525409\_2\_vis.fits} & 2000-12-14 & 22:38:39 & 1.28 & HI-RES & 2.2 \\
\texttt{C1355530771\_2\_vis.fits} & 2000-12-15 & 00:08:01 & 1.28 & HI-RES & 2.3 \\
\texttt{C1355536132\_2\_vis.fits} & 2000-12-15 & 01:37:22 & 1.28 & HI-RES & 2.5 \\
\texttt{C1355689391\_3\_vis.fits} & 2000-12-16 & 20:11:41 & 0.64 & NOMINAL & 5.8 \\
\texttt{C1355692689\_2\_vis.fits} & 2000-12-16 & 21:06:39 & 0.64 & NOMINAL & 5.9 \\
\texttt{C1355695987\_2\_vis.fits} & 2000-12-16 & 22:01:37 & 0.64 & NOMINAL & 5.9 \\
\texttt{C1355699285\_2\_vis.fits} & 2000-12-16 & 22:56:35 & 0.64 & NOMINAL & 6.0 \\
\texttt{C1355707028\_2\_vis.fits} & 2000-12-17 & 01:05:38 & 0.64 & NOMINAL & 6.2 \\
\texttt{C1355710326\_2\_vis.fits} & 2000-12-17 & 02:00:36 & 0.64 & NOMINAL & 6.3 \\
\texttt{C1355713623\_2\_vis.fits} & 2000-12-17 & 02:55:33 & 0.64 & NOMINAL & 6.4 \\
\texttt{C1355716921\_2\_vis.fits} & 2000-12-17 & 03:50:31 & 0.64 & NOMINAL & 6.5 \\
\texttt{C1355720219\_2\_vis.fits} & 2000-12-17 & 04:45:29 & 0.64 & NOMINAL & 6.5 \\
\texttt{C1355723517\_2\_vis.fits} & 2000-12-17 & 05:40:27 & 0.64 & NOMINAL & 6.6 \\
\texttt{C1355726815\_2\_vis.fits} & 2000-12-17 & 06:35:25 & 0.64 & NOMINAL & 6.7 \\
\texttt{C1356752365\_2\_vis.fits} & 2000-12-29 & 03:27:47 & 1.28 & HI-RES & 52.7 \\
\texttt{C1356756765\_2\_vis.fits} & 2000-12-29 & 04:41:07 & 1.28 & HI-RES & 53.0 \\
\texttt{C1356761165\_2\_vis.fits} & 2000-12-29 & 05:54:27 & 1.28 & HI-RES & 53.3 \\
\texttt{C1356976257\_3\_vis.fits} & 2000-12-31 & 17:39:17 & 1.28 & HI-RES & 68.0 \\
\texttt{C1356977354\_3\_vis.fits} & 2000-12-31 & 17:57:34 & 1.28 & HI-RES & 67.7 \\
\texttt{C1356980729\_1\_vis.fits} & 2000-12-31 & 18:53:50 & 0.64 & NOMINAL & 68.2 \\
\texttt{C1356989673\_1\_vis.fits} & 2000-12-31 & 21:22:54 & 0.64 & NOMINAL & 68.8 \\
\texttt{C1357108362\_2\_vis.fits} & 2001-01-02 & 06:21:02 & 0.64 & NOMINAL & 76.3 \\
\texttt{C1357116132\_1\_vis.fits} & 2001-01-02 & 08:30:31 & 1.28 & HI-RES & 76.8 \\
\texttt{C1357119162\_1\_vis.fits} & 2001-01-02 & 09:21:02 & 0.64 & NOMINAL & 77.0 \\
\texttt{C1357121532\_1\_vis.fits} & 2001-01-02 & 10:00:31 & 1.28 & HI-RES & 77.2 \\
\texttt{C1357558365\_2\_vis.fits} & 2001-01-07 & 11:21:01 & 1.28 & HI-RES & 99.7 \\
\texttt{C1357560735\_2\_vis.fits} & 2001-01-07 & 12:00:31 & 1.28 & HI-RES & 99.8 \\
\texttt{C1357563765\_2\_vis.fits} & 2001-01-07 & 12:51:01 & 1.28 & HI-RES & 99.9 \\
\texttt{C1357566135\_2\_vis.fits} & 2001-01-07 & 13:30:31 & 1.28 & HI-RES & 100.0 \\
\texttt{C1357569165\_2\_vis.fits} & 2001-01-07 & 14:21:01 & 1.28 & HI-RES & 100.1 \\
\texttt{C1357626912\_2\_vis.fits} & 2001-01-08 & 06:23:28 & 1.28 & HI-RES & 102.3 \\
\texttt{C1357627771\_2\_vis.fits} & 2001-01-08 & 06:37:47 & 1.28 & HI-RES & 102.4 \\
\texttt{C1357636771\_2\_vis.fits} & 2001-01-08 & 09:07:47 & 1.28 & HI-RES & 102.7 \\
\texttt{C1357644912\_2\_vis.fits} & 2001-01-08 & 11:23:28 & 1.28 & HI-RES & 103.0 \\
\texttt{C1357653912\_2\_vis.fits} & 2001-01-08 & 13:53:28 & 1.28 & HI-RES & 103.4 \\
\texttt{C1353677179\_2\_ir.fits} & 2000-11-23 & 13:15:03 & 0.02 & NOMINAL & 15.2 \\
\texttt{C1353694824\_2\_ir.fits} & 2000-11-23 & 18:09:08 & 0.02 & NOMINAL & 15.1 \\
\texttt{C1353820700\_2\_ir.fits} & 2000-11-25 & 05:07:03 & 0.02 & NOMINAL & 14.6 \\
\texttt{C1353838345\_2\_ir.fits} & 2000-11-25 & 10:01:08 & 0.02 & NOMINAL & 14.5 \\
\texttt{C1354107982\_2\_ir.fits} & 2000-11-28 & 12:55:03 & 0.02 & NOMINAL & 13.2 \\
\texttt{C1354125627\_2\_ir.fits} & 2000-11-28 & 17:49:08 & 0.02 & NOMINAL & 13.1 \\
\texttt{C1354251503\_2\_ir.fits} & 2000-11-30 & 04:47:03 & 0.02 & NOMINAL & 12.3 \\
\texttt{C1354269148\_2\_ir.fits} & 2000-11-30 & 09:41:08 & 0.02 & NOMINAL & 12.2 \\
\texttt{C1354538785\_2\_ir.fits} & 2000-12-03 & 12:35:03 & 0.02 & NOMINAL & 10.3 \\
\texttt{C1354556430\_2\_ir.fits} & 2000-12-03 & 17:29:08 & 0.02 & NOMINAL & 10.2 \\
\texttt{C1354682306\_2\_ir.fits} & 2000-12-05 & 04:27:03 & 0.02 & NOMINAL & 9.1 \\
\texttt{C1354969588\_2\_ir.fits} & 2000-12-08 & 12:15:03 & 0.02 & NOMINAL & 6.3 \\
\texttt{C1354987233\_2\_ir.fits} & 2000-12-08 & 17:09:08 & 0.02 & NOMINAL & 6.1 \\
\texttt{C1355233813\_3\_ir.fits} & 2000-12-11 & 13:38:46 & 0.02 & NOMINAL & 2.9 \\
\texttt{C1355237196\_3\_ir.fits} & 2000-12-11 & 14:35:09 & 0.02 & NOMINAL & 2.8 \\
\texttt{C1355240579\_3\_ir.fits} & 2000-12-11 & 15:31:32 & 0.02 & NOMINAL & 2.7 \\
\texttt{C1355243962\_3\_ir.fits} & 2000-12-11 & 16:27:55 & 0.02 & NOMINAL & 2.6 \\
\texttt{C1355247345\_3\_ir.fits} & 2000-12-11 & 17:24:18 & 0.02 & NOMINAL & 2.6 \\
\texttt{C1355252743\_3\_ir.fits} & 2000-12-11 & 18:54:16 & 0.02 & NOMINAL & 2.6 \\
\texttt{C1355256529\_3\_ir.fits} & 2000-12-11 & 19:57:22 & 0.02 & NOMINAL & 2.5 \\
\texttt{C1355259911\_3\_ir.fits} & 2000-12-11 & 20:53:44 & 0.02 & NOMINAL & 2.5 \\
\texttt{C1355263294\_3\_ir.fits} & 2000-12-11 & 21:50:07 & 0.02 & NOMINAL & 2.3 \\
\texttt{C1355266677\_3\_ir.fits} & 2000-12-11 & 22:46:30 & 0.02 & NOMINAL & 2.3 \\
\texttt{C1355309532\_3\_ir.fits} & 2000-12-12 & 10:40:45 & 0.02 & NOMINAL & 1.7 \\
\texttt{C1355313318\_3\_ir.fits} & 2000-12-12 & 11:43:51 & 0.02 & NOMINAL & 1.7 \\
\texttt{C1355316701\_3\_ir.fits} & 2000-12-12 & 12:40:14 & 0.02 & NOMINAL & 1.6 \\
\texttt{C1355320083\_3\_ir.fits} & 2000-12-12 & 13:36:36 & 0.02 & NOMINAL & 1.5 \\
\texttt{C1355323466\_3\_ir.fits} & 2000-12-12 & 14:32:58 & 0.02 & NOMINAL & 1.4 \\
\texttt{C1355326849\_3\_ir.fits} & 2000-12-12 & 15:29:21 & 0.02 & NOMINAL & 1.4 \\
\texttt{C1355330232\_3\_ir.fits} & 2000-12-12 & 16:25:44 & 0.02 & NOMINAL & 1.3 \\
\texttt{C1355333615\_3\_ir.fits} & 2000-12-12 & 17:22:07 & 0.02 & NOMINAL & 1.3 \\
\texttt{C1355336998\_3\_ir.fits} & 2000-12-12 & 18:18:30 & 0.02 & NOMINAL & 1.2 \\
\texttt{C1355340381\_3\_ir.fits} & 2000-12-12 & 19:14:53 & 0.02 & NOMINAL & 1.2 \\
\texttt{C1355518209\_2\_ir.fits} & 2000-12-14 & 20:38:39 & 0.02 & NOMINAL & 2.1 \\
\texttt{C1355521809\_2\_ir.fits} & 2000-12-14 & 21:38:40 & 0.02 & NOMINAL & 2.2 \\
\texttt{C1355525409\_2\_ir.fits} & 2000-12-14 & 22:38:39 & 0.02 & NOMINAL & 2.2 \\
\texttt{C1355529009\_2\_ir.fits} & 2000-12-14 & 23:38:40 & 0.02 & NOMINAL & 2.3 \\
\texttt{C1355530771\_2\_ir.fits} & 2000-12-15 & 00:08:01 & 0.02 & NOMINAL & 2.3 \\
\texttt{C1355534370\_2\_ir.fits} & 2000-12-15 & 01:08:01 & 0.02 & NOMINAL & 2.4 \\
\texttt{C1355536132\_2\_ir.fits} & 2000-12-15 & 01:37:22 & 0.02 & NOMINAL & 2.4 \\
\texttt{C1355539732\_2\_ir.fits} & 2000-12-15 & 02:37:23 & 0.02 & NOMINAL & 2.5 \\
\texttt{C1356752365\_2\_ir.fits} & 2000-12-29 & 03:27:47 & 0.02 & NOMINAL & 52.9 \\
\texttt{C1356756765\_2\_ir.fits} & 2000-12-29 & 04:41:07 & 0.02 & NOMINAL & 53.2 \\
\texttt{C1356761165\_2\_ir.fits} & 2000-12-29 & 05:54:27 & 0.02 & NOMINAL & 53.5 \\
\texttt{C1356907330\_3\_ir.fits} & 2000-12-30 & 22:30:31 & 0.02 & NOMINAL & 63.3 \\
\texttt{C1356910360\_3\_ir.fits} & 2000-12-30 & 23:21:01 & 0.02 & NOMINAL & 63.5 \\
\texttt{C1356911940\_3\_ir.fits} & 2000-12-30 & 23:47:21 & 0.02 & NOMINAL & 63.6 \\
\texttt{C1356913520\_3\_ir.fits} & 2000-12-31 & 00:13:41 & 0.02 & NOMINAL & 63.7 \\
\texttt{C1356976257\_3\_ir.fits} & 2000-12-31 & 17:39:17 & 0.02 & NOMINAL & 67.8 \\
\texttt{C1356977354\_3\_ir.fits} & 2000-12-31 & 17:57:34 & 0.02 & NOMINAL & 67.9 \\
\texttt{C1356980729\_1\_ir.fits} & 2000-12-31 & 18:53:50 & 0.02 & NOMINAL & 68.2 \\
\texttt{C1356989673\_1\_ir.fits} & 2000-12-31 & 21:22:54 & 0.02 & NOMINAL & 68.8 \\
\texttt{C1357108362\_2\_ir.fits} & 2001-01-02 & 06:21:02 & 0.02 & NOMINAL & 76.3 \\
\texttt{C1357110732\_2\_ir.fits} & 2001-01-02 & 07:00:31 & 0.02 & NOMINAL & 76.4 \\
\texttt{C1357113762\_2\_ir.fits} & 2001-01-02 & 07:51:02 & 0.02 & NOMINAL & 76.6 \\
\texttt{C1357119162\_1\_ir.fits} & 2001-01-02 & 09:21:02 & 0.02 & NOMINAL & 76.9 \\
\texttt{C1357312859\_1\_ir.fits} & 2001-01-04 & 15:09:18 & 0.02 & NOMINAL & 88.1 \\
\texttt{C1357326275\_1\_ir.fits} & 2001-01-04 & 18:52:54 & 0.02 & NOMINAL & 88.8 \\
\texttt{C1357330747\_1\_ir.fits} & 2001-01-04 & 20:07:26 & 0.02 & NOMINAL & 89.0 \\
\texttt{C1357335218\_1\_ir.fits} & 2001-01-04 & 21:21:57 & 0.02 & NOMINAL & 89.3 \\
\texttt{C1357758362\_2\_ir.fits} & 2001-01-09 & 18:54:18 & 0.02 & NOMINAL & 107.0 \\
\texttt{C1357767306\_1\_ir.fits} & 2001-01-09 & 21:23:22 & 0.02 & NOMINAL & 107.3 \\
\texttt{C1357771778\_1\_ir.fits} & 2001-01-09 & 22:37:54 & 0.02 & NOMINAL & 107.4 \\
\texttt{C1357776250\_1\_ir.fits} & 2001-01-09 & 23:52:26 & 0.02 & NOMINAL & 107.6 \\
\texttt{C1357780721\_1\_ir.fits} & 2001-01-10 & 01:06:57 & 0.02 & NOMINAL & 107.7 \\
\texttt{C1357861863\_1\_ir.fits} & 2001-01-10 & 23:39:18 & 0.02 & NOMINAL & 110.2 \\
\texttt{C1357866335\_1\_ir.fits} & 2001-01-11 & 00:53:50 & 0.02 & NOMINAL & 110.3 \\
\texttt{C1357870807\_1\_ir.fits} & 2001-01-11 & 02:08:22 & 0.02 & NOMINAL & 110.4 \\
\texttt{C1357875279\_1\_ir.fits} & 2001-01-11 & 03:22:54 & 0.02 & NOMINAL & 110.6 \\
\texttt{C1357879751\_1\_ir.fits} & 2001-01-11 & 04:37:26 & 0.02 & NOMINAL & 110.7 \\
\texttt{C1357884223\_1\_ir.fits} & 2001-01-11 & 05:51:58 & 0.02 & NOMINAL & 110.8 \\
\enddata
\end{deluxetable*}

\section{Dark Signal Correction Strategies}
\label{appendix:dark_correction}

This appendix summarizes the dark signal correction strategies considered for the VIS and IR channels, the behavior of the default ISIS/\texttt{vimscal} treatments on the Jupiter data set analyzed here, and the rationale for the corrections finally adopted in this work.

\subsection{VIS Channel}

\subsubsection{Default VIS dark correction in \texttt{vimscal}}

In the ISIS workflow, the VIS dark correction implemented in \texttt{vimscal} is stored in two calibration tables, one for NOMINAL sampling and one for HI-RES sampling (\texttt{vis\_lowres\_dark\_model\_v0001.tab} and \texttt{vis\_hires\_dark\_model\_v0001.tab}). The model provides two coefficients for each $(x,\lambda)$ pair, usually denoted $a(x,\lambda)$ and $b(x,\lambda)$, which are combined with the VIS exposure time to estimate the dark signal:
\begin{equation}
D(x,\lambda) = a(x,\lambda) + b(x,\lambda)\,t_{\mathrm{exp}},
\end{equation}
where $x$ is the detector sample and $\lambda$ is the VIS spectral band. The correction is then applied to each raw frame as
\begin{equation}
\mathrm{DN}_{\mathrm{corr}}(x,z,\lambda)=\mathrm{DN}(x,z,\lambda)-D(x,\lambda),
\end{equation}
i.e., it is constant along the line direction and varies only with sample and band. Separate models are used for NOMINAL and HI-RES sampling, reflecting the different detector sampling and binning schemes and offsets. The dark model is mapped onto the image frame using the detector-offset metadata, so that the appropriate segment of the full sample grid is applied to the cube.

\subsubsection{Empirical behavior in Jupiter VIS cubes and rationale for not adopting the default correction}

We tested the default ISIS VIS dark-model correction on the Jupiter data set used in this work and found that its behavior is strongly sampling-mode dependent and introduces undesirable artifacts for quantitative analysis.

In the NOMINAL cubes, applying the ISIS dark model produces an apparent extra subtraction: the off-disk background develops dark (low-value) stripes, and the overall background level becomes significantly negative, typically ranging from $-0.02$ to $-0.2$ in I/F depending on the spectral band. However, the most important point is that, under low-illumination conditions (i.e., near the limb or terminator), the dark signal correction leads to negative I/F values within the disk in dark spectral regions, such as the methane band near 890~nm, with the largest excursions reaching about $-0.02$ in I/F. 

In HI-RES, the VIS dark correction has no effect because, as shown by inspection of the \texttt{vimscal} source code, the implementation reads only the first 64 samples of the VIS dark-model table (the nominal-range indexing), whereas the HI-RES detector sample grid spans 192 samples. For the HI-RES cubes in our data set, the extracted image window maps to dark-model indices outside this 1--64 range; as a result, no dark values are retrieved and no dark signal subtraction is applied. We tested applying the HI-RES dark correction using a corrected mapping of the HI-RES dark model onto the extracted window. The outcome is worse than for nominal: although the HI-RES dark level is comparable to the nominal case ($\sim$50~DN), the measured disk signal in HI-RES is systematically lower than in nominal, so subtracting the dark model drives both the off-disk background and low-signal regions within the disk to significantly more negative $I/F$ values. In particular, negative $I/F$ values occur systematically within the disk in dark spectral regions, including pixels well away from the limb/terminator and extending to the disk center.

Taken together, these results support not applying any additional VIS dark signal subtraction to our raw Jupiter cubes. First, the VIS dark models used by \texttt{vimscal} have levels of order $\sim$50~DN across essentially the full spectral range, so their effect is not confined to a small subset of bands. Secondly, the raw VIS cubes already show a slightly negative off-disk background ($\sim$$-2$~DN on average), in contrast to the raw IR cubes, for which the dark correction is known to have been applied on board and whose background is much closer to zero ($\sim$$-0.2$~DN). Applying the VIS dark model therefore increases, rather than reduces, this discrepancy. In addition, the VIS cubes include a \texttt{SideplaneVis} table that stores tabulated values as a function of sample and band. These values are typically of order $\sim$50~DN, with relatively little variation across most bands, although valid entries are not present for all bands. By analogy with the IR channel, where the corresponding \texttt{SideplaneIr} table contains the dark values subtracted on board from the images, the \texttt{SideplaneVis} entries are likely to represent a dark-related instrumental signal. However, we found no clear evidence, either in the available documentation or in the cubes themselves, that the raw VIS data were dark-corrected using these stored values. Empirically, the most robust result is that applying an extra dark subtraction leads to over-correction, whereas leaving the VIS data without any additional dark correction yields very low signals in weak bands and under low-illumination geometries, but keeps them non-negative. We therefore do not apply any further dark correction to the VIS channel.

\subsubsection{Our VIS correction: stripe removal}

Across both VIS sampling modes, we identified a recurring set of bands affected by persistent stripe artifacts. In the data set analyzed here, these correspond to the 1-based band numbers 8, 10, 14, 17, 18, 25, 27, 58, and 94. In each affected band, the artifact is confined to a single detector sample, and the corresponding sample index is fixed within a given sampling mode.

We therefore correct only these empirically identified affected samples, treating HI-RES and NOMINAL cubes separately. Importantly, this correction is applied to the raw VIS cubes in DN space prior to conversion to I/F. As a result, the faintest disk pixels in the darkest VIS bands remain of order $\sim10^{-3}$ in I/F and stay non-negative.

\begin{figure}[ht!]
\centering
\includegraphics[width=0.7\linewidth]{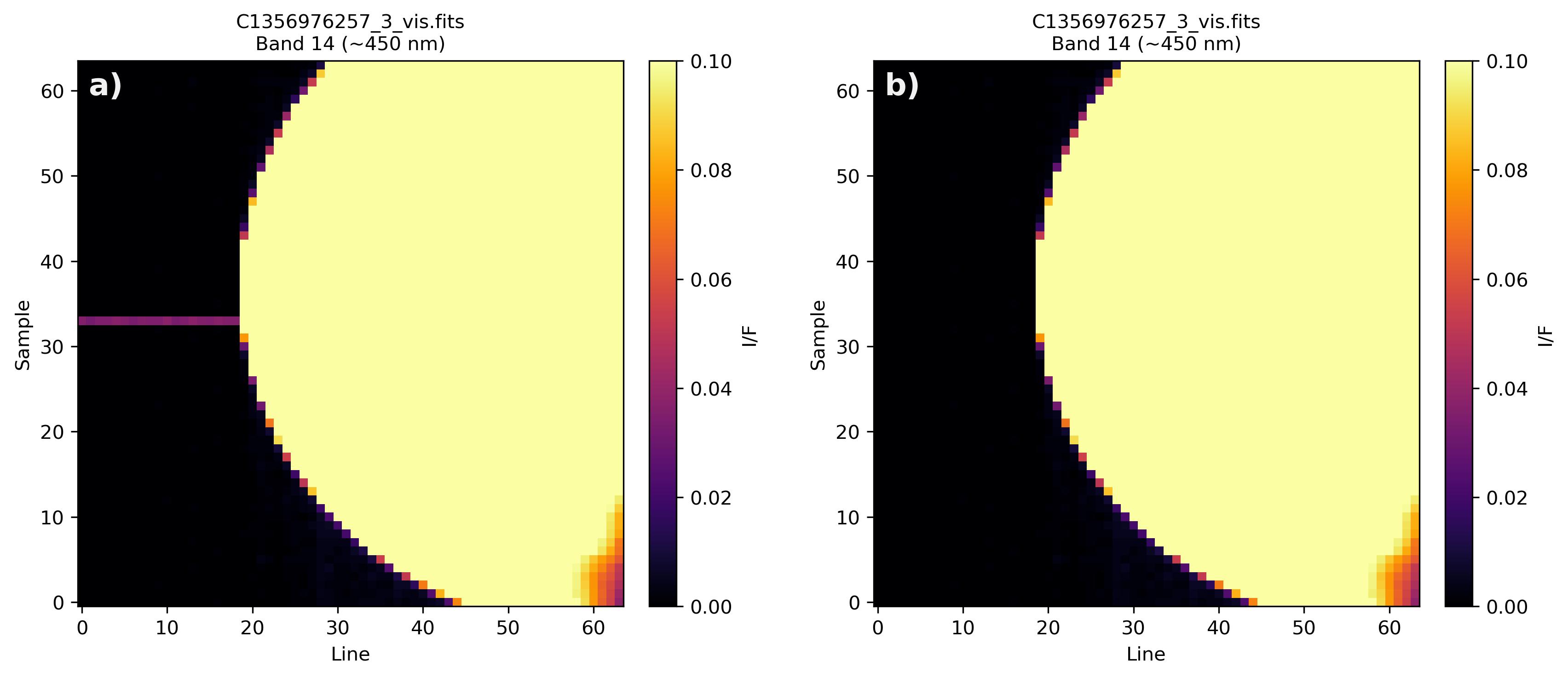}
\caption{Example of VIS cube \texttt{C1356976257\_3\_vis.fits}: (a) before and (b) after applying the adopted stripe-removal strategy. The image stretch has been adjusted to highlight the stripe artifact.}
\label{fig:vis_stripes}
\end{figure}

For each sampling mode and each affected band, we estimate the stripe amplitude as an additive offset relative to the local background. To obtain a robust estimate, this offset is derived from ten cubes in each sampling mode and then applied to all cubes of that mode. For each cube used in the estimate, the background level is measured from a sky patch in the image corners, using a $5\times5$ pixel box for NOMINAL cubes and a $10\times10$ pixel box for HI-RES cubes; the four corners are evaluated and the darkest is selected. The stripe amplitude for that cube is then computed as the difference between the disk-masked mean DN of the affected sample and the disk-masked background level. The final stripe offset adopted for a given sampling mode and band is the mean of these per-cube estimates, and its standard deviation provides a measure of the scatter. The adopted values are listed in Table~\ref{tab:vis_stripes}. This offset is then subtracted from the full affected sample for that band, thereby removing the stripe artifact, as illustrated in Figure~\ref{fig:vis_stripes}.

\begin{deluxetable}{cccccc}
\tabletypesize{\scriptsize}
\tablewidth{0pt}
\tablecaption{Mean VIS stripe deviations measured from the first ten raw cubes in each sampling mode. Band and sample indices are given in 1-based indexing.\label{tab:vis_stripes}}
\tablehead{
\colhead{Mode} &
\colhead{Band (1-based)} &
\colhead{Sample (1-based)} &
\colhead{Cubes used} &
\colhead{Stripe dev (DN)} &
\colhead{Stripe dev std (DN)}
}
\startdata
NOMINAL & 8  & 8  & 10 & 8.7  & 0.6 \\
NOMINAL & 10  & 12 & 10 & 12.2 & 0.6 \\
NOMINAL & 14 & 12 & 10 & 19.2 & 0.5 \\
NOMINAL & 17 & 5  & 10 & 6.5  & 1.0 \\
NOMINAL & 18 & 11 & 10 & 5.3  & 0.6 \\
NOMINAL & 25 & 12 & 10 & 16.6 & 0.6 \\
NOMINAL & 27 & 5  & 10 & 9.1  & 1.2 \\
NOMINAL & 58 & 22 & 10 & 10.8 & 1.2 \\
NOMINAL & 94 & 13 & 10 & 4.6  & 1.2 \\
\hline
HI-RES  & 8  & 21 & 10 & 14.5 & 3.0 \\
HI-RES  & 10  & 32 & 10 & 24.9 & 1.2 \\
HI-RES  & 14 & 31 & 10 & 37.9 & 1.3 \\
HI-RES  & 17 & 10  & 10 & 11.2 & 0.4 \\
HI-RES  & 18 & 28 & 10 & 9.9  & 0.5 \\
HI-RES  & 25 & 32 & 10 & 33.6 & 2.1 \\
HI-RES  & 27 & 11 & 10 & 17.8 & 2.0 \\
HI-RES  & 58 & 62 & 10 & 19.4 & 1.2 \\
HI-RES  & 94 & 35 & 10 & 6.1  & 0.5 \\
\enddata
\end{deluxetable}

\subsection{IR Channel}

\subsubsection{Default IR dark correction in \texttt{vimscal}}

In the ISIS workflow, the IR dark correction is based on the \texttt{SideplaneIr} table stored in the raw cube label. The sideplane provides one dark value for each $(z,\lambda)$ pair, where $z$ is the detector line and $\lambda$ is the IR spectral band. The sideplane therefore depends on detector line and wavelength.

The IR raw cubes used in this work already had the \texttt{SideplaneIr} values subtracted on board and are therefore dark-corrected. However, the sideplanes often contain anomalous spikes that, if used directly, produce dark stripes in the images. For this reason, \texttt{vimscal} applies by default a first-order polynomial fit to the sideplane values as a function of detector line and replaces the original sideplane correction with this fitted profile. Since the dark signal has already been subtracted on board, this is implemented as a \textit{Fit Delta} correction (in ISIS terminology): the pipeline subtracts $(\mathrm{fit}-\mathrm{sideplane})$ from data already expressed in $(\mathrm{DN}-\mathrm{sideplane})$ units, which is equivalent to replacing the original sideplane subtraction by subtraction of the fitted profile alone.

\subsubsection{Empirical behavior in Jupiter IR cubes and rationale for not adopting the default correction}

Inspection of the \texttt{SideplaneIr} tables in our Jupiter IR data set shows that the dark signal is generally close to constant with detector line for a given band, but is occasionally affected by isolated anomalous spikes. Because the raw IR cubes have already had the sideplane dark removed on board, such spikes appear in the images as over-subtracted dark stripes at the corresponding detector lines (see Figure~\ref{fig:ir_dark_correction_example}).

Applying the default ISIS \textit{Fit Delta} correction does not solve this problem satisfactorily. Since the line fit is performed using all valid sideplane values, anomalous spikes can bias the fitted profile and introduce an artificial line dependence into the correction. As a result, the default correction may spread the effect of a small number of anomalous sideplane values over the full line direction, rather than simply removing the localized artifact. This behavior is illustrated in Figure~\ref{fig:ir_dark_correction_example}.

\subsubsection{Our IR correction: sideplane replacement by a robust constant dark level}

For the IR cubes used in this work, we therefore do not adopt the default ISIS \textit{Fit Delta} correction. Instead, we replace the line-dependent sideplane correction by a robust constant dark level estimated independently for each spectral band. This correction is applied to the raw IR cubes in DN space prior to radiometric calibration.

Let $S(z,\lambda)$ denote the original sideplane value at detector line $z$ and wavelength $\lambda$. For each band, we first estimate a reference dark level from the mode of the sideplane values along the line direction. We then define inliers as those values satisfying

\begin{equation}
\left|S(z,\lambda)-\mathrm{mode}(\lambda)\right| \leq 20~\mathrm{DN}.
\end{equation}

\begin{figure}[ht!]
\centering
\includegraphics[width=0.55\columnwidth]{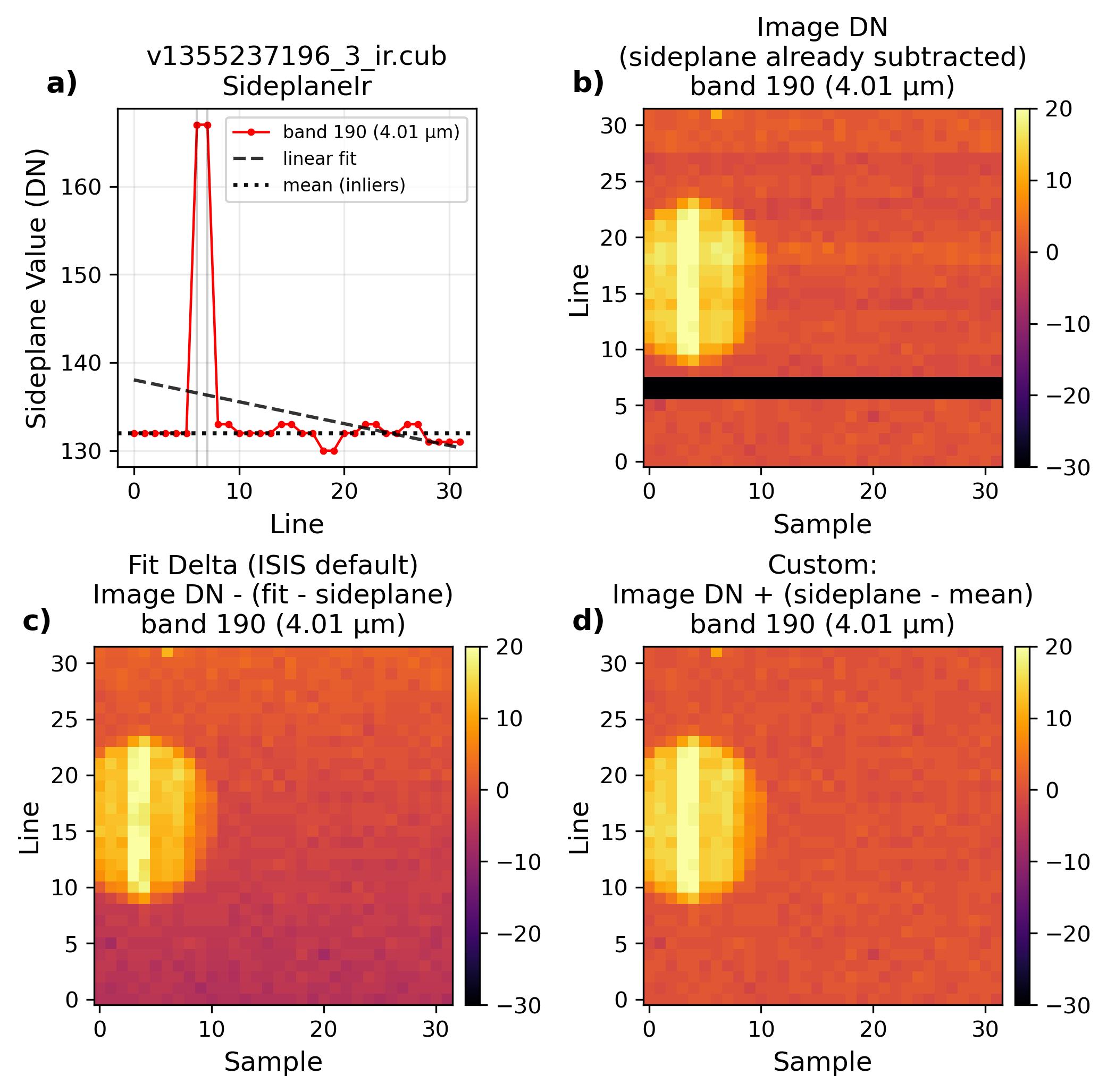}
\caption{Example of raw IR cube \texttt{v1355237196\_3\_ir.cub}, illustrating the adopted dark-correction strategy. (a) \texttt{SideplaneIr} values as a function of detector line for a band showing an anomalous spike, together with the first-order fit used by the default ISIS \textit{Fit Delta} approach and the mean value adopted here after excluding outliers. (b) Original raw image in DN after on-board sideplane subtraction, where the anomalous sideplane value produces an over-subtracted dark stripe. (c) Result obtained with the default ISIS \textit{Fit Delta} correction, which introduces an artificial large-scale gradient. (d) Result obtained with the adopted custom correction, in which the original sideplane subtraction is replaced by a robust constant dark level estimated from the mean of the sideplane inliers.}
\label{fig:ir_dark_correction_example}
\end{figure}

The adopted dark level for that band is taken to be the mean of these inlier values, $\bar{S}_{\rm in}(\lambda)$, which provides a robust estimate of the approximately constant dark signal while excluding isolated anomalous spikes. The corrected IR counts are then

\begin{equation}
\mathrm{DN}_{\rm corr}(x,z,\lambda)=\mathrm{DN}(x,z,\lambda)+S(z,\lambda)-\bar{S}_{\rm in}(\lambda),
\end{equation}

where $\mathrm{DN}(x,z,\lambda)$ are the raw IR counts after the on-board sideplane subtraction. In other words, we first undo the original line-dependent sideplane subtraction and then subtract a single robust band-dependent dark level. This preserves the overall dark offset of each band while removing the spurious line-based structure introduced by anomalous sideplane spikes. An example of the effect of our custom correction, compared with the original sideplane subtraction and the default \textit{Fit Delta} strategy applied by ISIS, is shown in Figure~\ref{fig:ir_dark_correction_example}.

\newpage
\bibliography{references}{}
\bibliographystyle{aasjournalv7}



\end{document}